\documentclass{article}
\usepackage{graphicx,times}
\begin {document}
\pagenumbering{arabic}
\pagestyle {plain}
\frenchspacing
\parindent 1.0 cm
\parskip 0.6cm
\vspace*{1.0 cm}
\begin{flushleft}
{\bf\large
Energy levels, radiative rates, and electron impact excitation rates for transitions in Li-like ions with 12 $\le$ Z $\le$ 20} \vspace{0.5 cm}
\\{\sf Kanti M. Aggarwal} and {\sf Francis P. Keenan}\\ \vspace*{0.3 cm} 
Astrophysics Research Centre, School of Mathematics and Physics, Queen's University Belfast,\\Belfast BT7 1NN,
Northern Ireland, UK.\\  \vspace*{0.1 cm}

\vspace*{0.2 cm}  
{\bf ABSTRACT} \\ 

We report calculations of energy levels, radiative rates, and electron impact excitation rates for transitions in Li-like ions with 12 $\le$ Z $\le$ 20.  The {\sc grasp}
(general-purpose relativistic atomic structure package) is adopted for calculating energy levels and radiative rates, while for determining the collision strengths and
subsequently the excitation rates, the Dirac atomic R-matrix code ({\sc darc}) is used.  Oscillator strengths, radiative rates, and line strengths are reported for all E1, E2,
M1, and M2 transitions among the lowest 24 levels of the Li-like ions considered. Collision strengths have been averaged over a Maxwellian velocity distribution, and the
effective collision strengths obtained are reported over a wide temperature range up to 10$^{7.4}$ K. Additionally, lifetimes are also listed for all calculated levels of the
above ions. Finally, extensive comparisons are made with available results in the literature, as well as with our parallel calculations for all parameters with the Flexible
Atomic Code  ({\sc fac}) in order to assess the accuracy of the reported results.

------------------------------------------------------------------------------------------------------------------------------------------------ \\
\vspace*{0.2 cm}
{\bf Running Title}: {\em K. M. Aggarwal and F. P. Keenan / Atomic Data and Nuclear Data Tables xxx (2011) xxx-xxx}
\end{flushleft}
\newpage

\begin{flushleft}
{\bf Contents}

\begin{tabular}{llr}
1.   &  Introduction  .........................................              &   00 \\
2.   &  Energy levels  .........................................             &   00 \\
3.   &  Radiative rates  .........................................           &   00 \\
4.   &  Lifetimes  .........................................                 &   00 \\
5.   &  Collision strengths  ..............................   		     &   00 \\
6.   &  Effective collision strengths  ..............                        &   00 \\
7.   &  Conclusions  ....................................		     &   00 \\
     &  Acknowledgments  .............................                       &   00 \\
     &  References  ........................................		     &   00 \\

\end{tabular}
\end{flushleft}
\newpage
\begin{flushleft}
   Explanation of Tables \\ \vspace*{0.2 cm}   
   Tables\\ \vspace*{0.2 cm} 

\begin{tabular}{rlr}
                                                                                                                     
 1a. & Energy levels of Mg X, their threshold energies (in Ryd), and lifetimes (in s) .................................................         &    00 \\
 1b. & Energy levels of Al XI, their threshold energies (in Ryd), and lifetimes (in s) .................................................        &    00 \\
 1c. & Energy levels of P XII, their threshold energies (in Ryd), and lifetimes (in s) .................................................        &    00 \\
 1d. & Energy levels of S XIV, their threshold energies (in Ryd), and lifetimes (in s) .................................................        &    00 \\
 1e. & Energy levels of Cl XV, their threshold energies (in Ryd), and lifetimes (in s) .................................................        &    00 \\
 1f. & Energy levels of Ar XVI, their threshold energies (in Ryd), and lifetimes (in s) .................................................       &    00 \\
 1g. & Energy levels of K XVII, their threshold energies (in Ryd), and lifetimes (in s) .................................................       &    00 \\
 1h. & Energy levels of Ca XVIII, their threshold energies (in Ryd), and lifetimes (in s) .................................................     &    00 \\
 2a. & Transition energies/wavelengths ($\lambda_{ij}$ in ${\rm \AA}$), radiative rates ($A_{ji}$ in s$^{-1}$), oscillator strengths .......... &    00 \\
     & (f$_{ij}$, dimensionless), line strengths S (in atomic unit) for electric dipole (E1), and ............................................. &    00 \\
     & A$_{ji}$ for electric quadrupole (E2), magnetic dipole (M1), and magnetic quadrupole (M2) transitions in Mg X. .....                     &    00 \\ 
 2b. & Transition energies/wavelengths ($\lambda_{ij}$ in ${\rm \AA}$), radiative rates ($A_{ji}$ in s$^{-1}$), oscillator strengths .......... &    00 \\
     & (f$_{ij}$, dimensionless), line strengths S (in atomic unit) for electric dipole (E1), and ............................................. &    00 \\
     & A$_{ji}$ for electric quadrupole (E2), magnetic dipole (M1), and magnetic quadrupole (M2) transitions in Al XI. .....                    &    00 \\  
 2c. & Transition energies/wavelengths ($\lambda_{ij}$ in ${\rm \AA}$), radiative rates ($A_{ji}$ in s$^{-1}$), oscillator strengths .......... &    00 \\
     & (f$_{ij}$, dimensionless), line strengths S (in atomic unit) for electric dipole (E1), and ............................................. &    00 \\
     & A$_{ji}$ for electric quadrupole (E2), magnetic dipole (M1), and magnetic quadrupole (M2) transitions in P XIII. ....                    &    00 \\  
 2d. & Transition energies/wavelengths ($\lambda_{ij}$ in ${\rm \AA}$), radiative rates ($A_{ji}$ in s$^{-1}$), oscillator strengths .......... &    00 \\
     & (f$_{ij}$, dimensionless), line strengths S (in atomic unit) for electric dipole (E1), and ............................................. &    00 \\
     & A$_{ji}$ for electric quadrupole (E2), magnetic dipole (M1), and magnetic quadrupole (M2) transitions in S XIV. ....                     &    00 \\  
 2e. & Transition energies/wavelengths ($\lambda_{ij}$ in ${\rm \AA}$), radiative rates ($A_{ji}$ in s$^{-1}$), oscillator strengths .......... &    00 \\
     & (f$_{ij}$, dimensionless), line strengths S (in atomic unit) for electric dipole (E1), and ............................................. &    00 \\
     & A$_{ji}$ for electric quadrupole (E2), magnetic dipole (M1), and magnetic quadrupole (M2) transitions in Cl XV. ....                     &    00 \\  
 2f. & Transition energies/wavelengths ($\lambda_{ij}$ in ${\rm \AA}$), radiative rates ($A_{ji}$ in s$^{-1}$), oscillator strengths .......... &    00 \\
     & (f$_{ij}$, dimensionless), line strengths S (in atomic unit) for electric dipole (E1), and ............................................. &    00 \\
     & A$_{ji}$ for electric quadrupole (E2), magnetic dipole (M1), and magnetic quadrupole (M2) transitions in Ar XVI. ....                    &    00 \\  
 2g. & Transition energies/wavelengths ($\lambda_{ij}$ in ${\rm \AA}$), radiative rates ($A_{ji}$ in s$^{-1}$), oscillator strengths .......... &    00 \\
     & (f$_{ij}$, dimensionless), line strengths S (in atomic unit) for electric dipole (E1), and ............................................. &    00 \\
     & A$_{ji}$ for electric quadrupole (E2), magnetic dipole (M1), and magnetic quadrupole (M2) transitions in K XVII. ....                    &    00 \\  
 2h. & Transition energies/wavelengths ($\lambda_{ij}$ in ${\rm \AA}$), radiative rates ($A_{ji}$ in s$^{-1}$), oscillator strengths .......... &    00 \\
     & (f$_{ij}$, dimensionless), line strengths S (in atomic unit) for electric dipole (E1), and ............................................. &    00 \\
     & A$_{ji}$ for electric quadrupole (E2), magnetic dipole (M1), and magnetic quadrupole (M2) transitions in Ca XVIII. ...                   &    00 \\  
 3a. & Collision strengths for resonance transitions in Mg X .................................................................................  &    00 \\
 3b. & Collision strengths for resonance transitions in Al XI ................................................................................  &    00 \\ 
 3c. & Collision strengths for resonance transitions in P XIII ...............................................................................  &    00 \\
 3d. & Collision strengths for resonance transitions in S XIV ................................................................................  &    00 \\
 3e. & Collision strengths for resonance transitions in Cl XV ................................................................................  &    00 \\
 3f. & Collision strengths for resonance transitions in Ar XVI ...............................................................................  &    00 \\
 3g. & Collision strengths for resonance transitions in K XVII ...............................................................................  &    00 \\		
 3h. & Collision strengths for resonance transitions in Ca XVIII .............................................................................  &    00 \\
 4a. & Effective collision strengths for transitions in Mg X .................................................................................  &    00 \\
 4b. & Effective collision strengths for transitions in Al XI ................................................................................  &    00 \\ 
 4c. & Effective collision strengths for transitions in P XIII ...............................................................................  &    00 \\ 
 4d. & Effective collision strengths for transitions in S XIV ................................................................................  &    00 \\ 
 4e. & Effective collision strengths for transitions in Cl XV ................................................................................  &    00 \\ 
 4f. & Effective collision strengths for transitions in Ar XVI ...............................................................................  &    00 \\ 
 4g. & Effective collision strengths for transitions in K XVII ...............................................................................  &    00 \\ 
 4h. & Effective collision strengths for transitions in Ca XVIII .............................................................................  &    00 \\ 
\end{tabular}	
\end{flushleft}															      

\newpage
\begin{flushleft}
{\bf 1. Introduction}
\end{flushleft}

\parindent = 1 cm 

Emission lines of Li-like ions have been observed in the spectra of a variety of astrophysical \cite{kpd}-\cite{df} and laboratory plasmas \cite{fbr}-\cite{ile}. Similarly, high
gain lengths have been measured in lasing plasmas, particularly for the 3d--4f transitions of Al XI and Si XII \cite{al11a} in the soft x-ray region. More recently, amplified
spontaneous emission signals of 3d--4d and 3d--4f transitions of Al XI have been observed by Yamaguchi et al. \cite{al11b}. In fact, 3d--4f (apart from other 2--3 and 3--4)
transitions of Li-like ions are of very high interest for x-ray lasers, as demonstrated by Brown et al. \cite{al11c} for ions of P, S, Cl, and K. Transitions of many Li-like
ions are also useful for temperature and density diagnostics of solar plasmas \cite{df}, \cite{ar16a}  as well as for the determination of chemical abundances \cite{ca20a}.
However to analyse observations, atomic data are required for a variety of parameters, such as energy levels, radiative rates (A-values), and excitation rates or equivalently
the effective collision strengths ($\Upsilon$), which are obtained from the electron impact collision strengths ($\Omega$). These data are also required for the  modelling of
fusion plasmas \cite{al11d} as many elements are present as impurities in the reactor walls. With this in view we have already reported calculations for Li-like ions with Z
$\le$ 11 \cite{ovi}-\cite{nv} and in this paper we report similar results for ions with 12 $\le$ Z $\le$ 20. It may also be noted that data for transitions of Si XII have
already been reported \cite{sixii} and are therefore not included in this paper.

Experimentally, energy levels for Li-like ions have been compiled by NIST (National Institute of Standards and Technology) and are available at their website {\tt
http://www.nist.gov/pml/data/asd.cfm}. Theoretically, energy levels  and A-values have been calculated by Nahar \cite{sn} for transitions of 15 Li-like ions in the 6 $\le$ Z
$\le$ 28 range. However, collisional atomic data are limited to transitions among the lowest three levels of Li-like ions with 8 $\le$ Z $\le$ 92, and from these levels to the
higher excited levels of the $n \le$ 5 configurations \cite{hlz}. The calculations of Zhang et al. \cite{hlz} are based on the relativistic {\em distorted-wave} (DW) method,
and  collision strengths ($\Omega$) are reported only at six energies above thresholds. Since resonances in the thresholds region have not been resolved, the subsequent results
for effective collision strengths ($\Upsilon$) are significantly underestimated over a wide temperature range for a large number of transitions (in particular forbidden ones),
as already demonstrated in our earlier papers for ions with Z $\le$ 11 \cite{ovi}-\cite{nv} and Si XII  \cite{sixii}. 

In this paper we report a complete set of results (namely energy levels, radiative rates, and effective collision strengths) for all transitions among the lowest 24 levels of
Li-like ions with 12 $\le$ Z $\le$ 20. Additionally, we also provide the A-values for four types of transitions, namely electric dipole (E1), electric quadrupole (E2), magnetic
dipole (M1), and  magnetic quadrupole (M2), because these are also required for  plasma modelling. For our calculations we employ the fully relativistic {\sc grasp}
(general-purpose relativistic atomic structure  package) code for the determination of wavefunctions,  originally developed by Grant et al. \cite{grasp0} and revised by Dr. 
P.H. Norrington. It is a fully relativistic code, and is based on the $jj$ coupling scheme.  Further relativistic corrections arising from the Breit interaction and QED effects
(vacuum polarization and Lamb shift) have also been included. In calculating energy levels, we have used the option of {\em extended average level} (EAL),  in which a weighted
(proportional to 2$j$+1) trace of the Hamiltonian matrix is minimized. This produces a compromise set of orbitals describing closely lying states with  moderate accuracy.
Similarly, for our calculations of $\Omega$, we have adopted the {\em Dirac atomic $R$-matrix code} ({\sc darc}) of P.H. Norrington and I.P. Grant (private communication).
Finally, to make comparisons and to assess the accuracy of our results, we have performed parallel calculations from the {\em Flexible Atomic Code} ({\sc fac}) of Gu \cite{fac},
available from the website {\tt {\verb+http://sprg.ssl.berkeley.edu/~mfgu/fac/+}}. This is also a fully relativistic code which provides a variety of atomic parameters, and
yields results comparable to {\sc grasp} and {\sc darc} - see, for example, Aggarwal et al. \cite{fe26b} and references therein. Generally, there is no major discrepancy in the
determination of energy levels and radiative rates from {\sc grasp} and {\sc fac}, but the former code allows more flexibility in terms of optimisations. Furthermore, the
identification of levels belonging to a particular configuration is a greater problem in {\sc fac} than in {\sc grasp}, although it should not apply to simple Li-like ions. On
the other hand, differences in collision strengths can sometimes be larger, particularly for the forbidden transitions, as discussed and demonstrated in some of our earlier
papers \cite{ovi}-\cite{sixii}. Similarly, results for $\Upsilon$ can be significantly underestimated by the {\sc fac} code, particularly for forbidden transitions and at lower
temperatures, due to the neglect of the contribution of resonances. Nevertheless, results from {\sc fac} will be helpful in assessing the accuracy of our energy levels,
radiative rates, and collision strengths, and in estimating the contribution of resonances for the determination of $\Upsilon$ values.

\begin{flushleft}
{\bf 2. Energy levels}
\end{flushleft}
\parindent = 1 cm 

The 1s$^2$ ${n}\ell$ (2 $\le$ $n \le$ 5) configurations of Li-like ions give rise to the lowest 24 energy levels listed in Tables 1 (a--h). Included in these tables are  our
level energies from {\sc grasp}, obtained {\em without} and {\em with} the inclusion of Breit and QED effects, plus the experimental energies compiled by NIST. Also included in
these tables are the theoretical energies obtained from our parallel calculations from {\sc fac} and those by Nahar \cite{sn} from the Breit-Pauli R-matrix (BPRM) code of
Berrington et al. \cite{rm1}.  The inclusion of the Breit and QED corrections decreases the level energies (except for level 2, i.e. 1s$^2$ $^2$P$^o_{1/2}$)  by a maximum of
$\sim$ 0.03 Ryd, depending on the ion. As the ion nuclear charge $Z$ increases, so does the effect of the Breit and QED corrections. For example, energy levels of Mg X decrease
by  $\sim$ 0.005 Ryd, but for Ca XVIII by $\sim$ 0.03 Ryd. Clearly, the importance of relativistic effects increases with increasing $Z$. Our Breit and QED corrected energies
from {\sc grasp} agree closely (generally within 0.01 Ryd) with those of NIST and the orderings are also nearly the same, although there are some noticeable differences, such as
for the 4d,5d $^2$D$_{3/2,5/2}$  levels of K XVII and Ca XVIII. Degeneracy for these levels in the NIST compilations is up to 0.061 Ryd, whereas it is $\sim$ 0.01 Ryd in all the
calculations listed in Tables 1 (a--h). Furthermore, the NIST energies for some of the levels are non-degenerate -- see, for example, the 4p,5p $^2$P$^o_{1/2,3/2}$  levels of Mg
X and  5d $^2$D$_{3/2,5/2}$  levels of P XIII. Similarly, NIST energies are not available for some of the levels -- see, for example,  the 5f $^2$F$^o_{5/2,7/2}$ and 5g
$^2$G$_{7/2,9/2}$ levels of P XIII, Cl XV, Ar XVI, K XVII, and Ca XVIII. 

The energies obtained from {\sc fac} are generally comparable with the experimental compilations of NIST and/or our calculations from {\sc grasp}. However, the orderings of some
of the levels are slightly different from {\sc fac} than from the experimental and other theoretical energies. As examples, see 12--13 and 19--24 levels of Mg X and 16--24 of Ca
XVIII. Nevertheless, energy differences among these levels are very small. Finally, the BPRM energies of Nahar \cite{sn} have the same orderings as NIST and our calculations
from {\sc grasp}. The BPRM energies also generally agree in magnitude, except for 4s,5s $^2$S$_{1/2}$ levels for which they are lower by up to 0.2 Ryd, depending on the ion.
Overall, we may state that there is no (major) discrepancy between theory and experiment for the energy levels of Li-like ions.

\begin{flushleft}
{\bf 3. Radiative rates}
\end{flushleft}

The absorption oscillator strength ($f_{ij}$) and radiative rate $A_{ji}$ (in s$^{-1}$) for a transition $i \to j$ are related by the following expression:

\begin{equation}
f_{ij} = \frac{mc}{8{\pi}^2{e^2}}{\lambda^2_{ji}} \frac{{\omega}_j}{{\omega}_i} A_{ji}
 = 1.49 \times 10^{-16} \lambda^2_{ji} (\omega_j/\omega_i) A_{ji}
\end{equation}
where $m$ and $e$ are the electron mass and charge, respectively, $c$ is the velocity of light,  $\lambda_{ji}$ is the transition energy/wavelength in $\rm \AA$, and $\omega_i$
and $\omega_j$ are the statistical weights of the lower ($i$) and upper ($j$) levels, respectively. Similarly, the oscillator strength $f_{ij}$ (dimensionless) and the line
strength $S$ (in atomic units, 1 a.u. = 6.460$\times$10$^{-36}$ cm$^2$ esu$^2$) are related by the  following standard equations:

\begin{flushleft}
for the electric dipole (E1) transitions 
\end{flushleft} 
\begin{equation}
A_{ji} = \frac{2.0261\times{10^{18}}}{{{\omega}_j}\lambda^3_{ji}} S^{{\rm E1}} \hspace*{0.5 cm} {\rm and} \hspace*{0.5 cm} 
f_{ij} = \frac{303.75}{\lambda_{ji}\omega_i} S^{{\rm E1}}, \\
\end{equation}
\begin{flushleft}
for the magnetic dipole (M1) transitions  
\end{flushleft}
\begin{equation}
A_{ji} = \frac{2.6974\times{10^{13}}}{{{\omega}_j}\lambda^3_{ji}} S^{{\rm M1}} \hspace*{0.5 cm} {\rm and} \hspace*{0.5 cm}
f_{ij} = \frac{4.044\times{10^{-3}}}{\lambda_{ji}\omega_i} S^{{\rm M1}}, \\
\end{equation}
\begin{flushleft}
for the electric quadrupole (E2) transitions 
\end{flushleft}
\begin{equation}
A_{ji} = \frac{1.1199\times{10^{18}}}{{{\omega}_j}\lambda^5_{ji}} S^{{\rm E2}} \hspace*{0.5 cm} {\rm and} \hspace*{0.5 cm}
f_{ij} = \frac{167.89}{\lambda^3_{ji}\omega_i} S^{{\rm E2}}, 
\end{equation}

\begin{flushleft}
and for the magnetic quadrupole (M2) transitions 
\end{flushleft}
\begin{equation}
A_{ji} = \frac{1.4910\times{10^{13}}}{{{\omega}_j}\lambda^5_{ji}} S^{{\rm M2}} \hspace*{0.5 cm} {\rm and} \hspace*{0.5 cm}
f_{ij} = \frac{2.236\times{10^{-3}}}{\lambda^3_{ji}\omega_i} S^{{\rm M2}}. \\
\end{equation}

In Tables 2 (a--h) we present transition energies/wavelengths ($\lambda$, in $\rm \AA$), radiative rates ($A_{ji}$, in s$^{-1}$), oscillator strengths ($f_{ij}$, dimensionless),
and line strengths ($S$, in a.u.), in length  form only, for all 92 electric dipole (E1) transitions among the 24 levels of Li-like ions with 12 $\le$ Z $\le$ 20. The {\em
indices} used  to represent the lower and upper levels of a transition have already been defined in Tables 1 (a--h). Similarly, there are 100 electric quadrupole (E2), 79 
magnetic dipole (M1), and 108 magnetic quadrupole (M2) transitions among the 24 levels. However, for these transitions only the A-values are listed in Tables 2 (a--h), and the
corresponding results for f- or S-values may be easily obtained using Eqs. (1--5).

In Tables A--H we compare our A-values from the {\sc grasp} and {\sc fac} codes for some of the common E1 transitions with those of Nahar \cite{sn} from the BPRM code of
Berrington et al. \cite{rm1}. Also included in these tables are the f-values from our {\sc grasp} calculations. In general, the A-values from the {\sc grasp} and {\sc fac} codes
agree satisfactorily for a majority of transitions, although there are differences of up to 30\% for a few, such as 1--17 and 1--18 of Mg X, for which there are no discrepancies
with the BPRM calculations of Nahar. However, these differences between two independent calculations decrease with increasing $Z$, i.e. reduce to 13\% for the corresponding
transitions of Ca XVIII. On the other hand,  there are significant differences with the Nahar calculations, of up to two orders of magnitude, particularly for the comparatively
weaker (f $\le$ 0.01) transitions, such as 7--11, 7--18, and 12--18. For these transitions the BPRM A-values of Nahar are {\em lower} for all Li-like ions, as also noted for
ions with Z $\le$ 11 \cite{nv}-\cite{sixii}. The A-values for such weak transitions from the BPRM code are likely to be less accurate, as discussed by Hibbert \cite{ah4} and
Aggarwal et al. \cite{fe17}. In addition, several transitions are missing from the Nahar calculations, and examples include:  2p $^2$P$^o_{1/2}$ -- 3s $^2$S$_{1/2}$ (2--4), 2p
$^2$P$^o_{3/2}$ -- 3s $^2$S$_{1/2}$ (3--4), and 2p $^2$P$^o_{3/2}$ -- 3d $^2$D$_{3/2}$ (3--7). Finally, there are differences of up to a factor of four between the BPRM 
A-values of Nahar and our calculations from the {\sc grasp} and {\sc fac} codes for several transitions, particularly those among levels 16 and higher. For all such transitions,
the energy differences ($\Delta E_{ij}$) are very small and hence small change in these affect the A-values significantly. 

One of the general criteria to assess the accuracy of radiative rates is to compare the length and velocity forms of the f- or A-values. However, such comparisons are only
desirable, but are {\em not} a fully sufficient test to assess accuracy, as different calculations (or combinations of configurations) may give comparable f-values in the two
forms, but entirely different results in magnitude. Generally, there is good agreement between the length and velocity forms of the f-values for {\em strong} transitions, but
differences  between the two forms can sometimes be substantial even for some very strong transitions, as demonstrated through various examples by Aggarwal et al. \cite{fe15c}.
However, for transitions of Li-like ions, there is no discrepancy between the two forms, and this is irrespective of the magnitude of a transition. Furthermore, to assess the
importance of configuration interaction (CI), we have performed another calculation including the levels of the $n$ = 6 configurations. Only for two transitions, namely 2--16 
(2p $^2$P$^o_{1/2}$ -- 5s $^2$S$_{1/2}$) and 3--16 (2p $^2$P$^o_{3/2}$ -- 5s $^2$S$_{1/2}$), have the A-values decreased by 30\% for Mg X and by 20\% for Ca XVIII. For the
remaining transitions the effect of additional CI is much less noticeable. Based on this and other comparisons discussed above, we may state that the A-values listed in Tables 2
(a--h) from our {\sc grasp} calculations are accurate to better than 20\% for all transitions and all ions, and hence should be the best currently available.

\begin{flushleft}
{\bf 4. Lifetimes}
\end{flushleft}

The lifetime $\tau$ for a level $j$ is defined as follows:

\begin{equation}  {\tau}_j = \frac{1}{{\sum_{i}^{}} A_{ji}}.  
\end{equation}  

Since this is a measurable parameter, it provides a check on the accuracy of the calculations. Therefore, in Tables 1 (a--h) we have also listed our calculated lifetimes from
the {\sc grasp} code, which {\em include} the contributions from four types of transitions, i.e. E1, E2, M1, and M2. To our knowledge, no measurements are available for
lifetimes in Li-like ions with 12 $\le$ Z $\le$ 20. However, Ishii et al. \cite{cl15a} have measured lifetimes from beam foil experiments for the 2p $^2$P$^o_{1/2,3/2}$ levels
of Cl XV, namely 1.0$\pm$0.1 and 0.71$\pm$0.03 ns, respectively. Both these measured values agree very well with our calculated lifetimes. Froese-Fischer has calculated
lifetimes for all the Li-like ions under discussion here and has posted her results for some levels up to 4s $^2$S$_{1/2}$ at her website: {\tt
{\verb+http://www.vuse.vanderbilt.edu/~cff/mchf_collection/+}}. There are no discrepancies for any of the levels between our and her calculations. We hope the present results
for other levels will be useful for future comparisons and may also encourage experimentalists to measure lifetimes for the Li-like  ions.

\begin{flushleft}
{\bf 5. Collision strengths}
\end{flushleft}

For the computation of collision strengths ($\Omega$) we have employed the {\em Dirac atomic $R$-matrix code} ({\sc darc}), which includes the relativistic effects in a
systematic way, in both the target description and the scattering model. It is based on the $jj$ coupling scheme, and uses the  Dirac-Coulomb Hamiltonian in the $R$-matrix
approach. The $R$-matrix radii adopted for Li-like ions with 12 $\le$ Z $\le$ 20 (except $Z$ = 14) are 9.60, 8.64, 7.20, 6.40, 5.92, 5.44, 5.28, and 5.12 au, respectively. For
all eight ions, 56 continuum orbitals have been included for each channel angular momentum for the expansion of the wavefunction. This allows us to compute $\Omega$ up to an
energy of  180, 215, 310, 390, 450, 540, 570,  and 600 Ryd, for 12 $\le$ Z $\le$ 20, respectively. Furthermore, these energy ranges are sufficient to calculate values of
$\Upsilon$ up to T$_e$ = 10$^{7.4}$ K, i.e. well above the  temperature of maximum abundance of these ions in ionisation equilibrium - see Bryans et al. \cite{pb}. The maximum
number of channels for a partial wave is 108, and the corresponding size of the Hamiltonian matrix is 6086. To obtain convergence of  $\Omega$ for all transitions and at all
energies, we have included all partial waves with angular momentum $J \le$ 60, although a larger number would have been  preferable for the convergence of some allowed
transitions, especially at higher energies. However, to account for neglected higher partial waves, we have included a top-up, based on the Coulomb-Bethe \cite{ab} approximation
for allowed transitions and geometric series for others.

For illustration, in Figs. 1-3 we show the variation of $\Omega$ with angular momentum $J$ for three transitions of Ar XVI, namely 1--3 (2s $^2$S$_{1/2}$ -- 2p
$^2$P$^o_{3/2}$),  1--8 (2s $^2$S$_{1/2}$ -- 3d $^2$D$_{5/2}$), and 1--10 (2s $^2$S$_{1/2}$ -- 4p $^2$P$^o_{1/2}$), respectively, and at five energies of 100, 200, 300, 400, and
500 Ryd. Values of $\Omega$ have fully converged for all forbidden (such as 1-8) and most of the allowed (such as 1--10) transitions. However, for some allowed transitions, such
as 1--3 shown in Fig. 1, our adopted range of partial waves ($J \le$ 60) is not sufficient for the convergence of $\Omega$ at most of the energies, particularly above
thresholds.  For such transitions a top-up has been included as mentioned above, and has been found to be appreciable. A similar conclusion applies to the transitions in other
Li-like ions.

In Tables 3 (a--h) we list our values of $\Omega$ for resonance transitions in Li-like ions with 12 $\le$ Z $\le$ 20 at energies {\em above} thresholds. The {\em indices} used 
to represent the levels of a transition have already been defined in Tables 1 (a--h). Very limited data are available for comparison, as already stated in section 1. Therefore,
to make an accuracy assessment of the values of $\Omega$, we have performed another calculation using the {\sc fac} code of Gu \cite{fac}. This code is also fully relativistic,
and is based on the well-known and widely-used {\em distorted-wave} (DW) method.  Furthermore, the same CI is included in {\sc fac} as in the calculations from {\sc darc}.
Therefore, also included in Tables 3 (a--h) for comparison purposes are the $\Omega$ values from {\sc fac} at a single {\em excited} energy E$_j$, which corresponds to $\sim$
140, 170, 230, 400, 450, 510, 580, and 450 Ryd for 12 $\le$ Z $\le$ 20, respectively. 

Generally the two sets of $\Omega$ from {\sc darc} and {\sc fac} agree to within 20\% for a majority of transitions. However for $\sim$ 15\% of the transitions there are
differences of  over 20\%, and for $\sim$ 5\% the discrepancies between the two independent calculations are over 50\%. Most of the transitions for which the discrepancies are
large are forbidden, and one of the main reasons is the anomalous behaviour of $\Omega$ in the calculations from the {\sc fac} code. To illustrate this, in Fig. 4 we compare the
two sets of $\Omega$ from {\sc darc} and {\sc fac} for three allowed transitions of Ar XVI, namely 4--6 (3s $^2$S$_{1/2}$ -- 3p $^2$P$^o_{3/2}$), 6--8 (3p $^2$P$^o_{3/2}$  -- 3d
$^2$D$_{5/2}$), and 10--12 (4p $^2$P$^o_{1/2}$ -- 4d $^2$D$_{3/2}$). For these three and many other allowed transitions there are no major discrepancies between the two sets of
$\Omega$, although there is a fall in magnitude for the 10--12 (and some other) transitions at the higher end of the energy range for {\em all} ions. In Fig. 5 we show similar
comparisons for three forbidden transitions, namely 6--11 (3p $^2$P$^o_{3/2}$  -- 4p $^2$P$^o_{3/2}$), 8--13 (3d $^2$D$_{5/2}$ -- 4d $^2$D$_{5/2}$), and 10--13 (4p
$^2$P$^o_{1/2}$ -- 4d $^2$D$_{5/2}$). Differences for these three (and some other) transitions are generally within 50\%, but are up to a factor of two for a few, such as
12--16, 13--16, 16--18, 19--21, and 20--22. For some transitions, such as 6--11, there are differences over the entire energy range, while for others such as 8--13, the
discrepancies are mainly at the lower end of the energy range, and completely disappear at large energies.  In some cases, such as 10--13, the differences between the two sets
of $\Omega$ are mainly at the higher end of the energy range due to a sudden anomaly in $\Omega$ from  the {\sc fac} code. Discrepancies in the values of $\Omega$ for some
transitions from {\sc darc} and {\sc fac} are partly due to the differences in the methodology, i.e. $R$-matrix and DW, respectively, and partly because {\sc fac} is designed to
generate a large amount of atomic data in a comparatively very short period of time, and without too much loss of accuracy. This aim is achieved by employing various
interpolation and extrapolation techniques, which result in significant differences in the values of $\Omega$ for some transitions and at some energies, as noted above. Overall,
we may state that for transitions for which there is no anomaly in the calculations of $\Omega$ from  the {\sc fac} code, the differences  between the {\sc darc} and {\sc fac}
calculations of collision strengths are generally within 20\%. However, the discrepancies are significant for those transitions for which values of $\Omega$ from  the {\sc fac}
code are anomalous. A similar conclusion applies to transitions in other Li-like ions. Finally, based on the comparisons made above as well as for other Li-like ions
\cite{ovi}-\cite{sixii}, we estimate our results for $\Omega$ to be accurate to better than 20\% for a majority of the transitions of all eight Li-like ions. 

\begin{flushleft}
{\bf 6. Effective collision strengths}
\end{flushleft}

Excitation rates, along with energy levels and radiative rates, are required for plasma modelling, and are determined from the collision strengths ($\Omega$). Since the
threshold energy region is dominated by numerous closed-channel (Feshbach) resonances, values of $\Omega$ need to be calculated in a fine energy mesh in order to accurately
account for their contribution. Furthermore, in a hot plasma electrons have a wide distribution of velocities, and therefore values of $\Omega$ are generally averaged over a
{\em Maxwellian} distribution as follows:

\begin{equation}
\Upsilon(T_e) = \int_{0}^{\infty} {\Omega}(E) {\rm exp}(-E_j/kT_e) d(E_j/{kT_e}),
\end{equation}
where $k$ is Boltzmann constant, T$_e$ is electron temperature in K, and E$_j$ is the electron energy with respect to the final (excited) state. Once the value of $\Upsilon$ is
known the corresponding results for the excitation $q(i,j)$ and de-excitation $q(j,i)$ rates can be easily obtained from the following equations:

\begin{equation}
q(i,j) = \frac{8.63 \times 10^{-6}}{{\omega_i}{T_e^{1/2}}} \Upsilon {\rm exp}(-E_{ij}/{kT_e}) \hspace*{1.0 cm}{\rm cm^3s^{-1}}
\end{equation}
and
\begin{equation}
q(j,i) = \frac{8.63 \times 10^{-6}}{{\omega_j}{T_e^{1/2}}} \Upsilon \hspace*{1.0 cm}{\rm cm^3 s^{-1}},
\end{equation}
where $\omega_i$ and $\omega_j$ are the statistical weights of the initial ($i$) and final ($j$) states, respectively, and $E_{ij}$ is the transition energy. The contribution of
resonances may enhance the values of $\Upsilon$ over those of the background values of collision strengths ($\Omega_B$), especially for the forbidden transitions, by up to a
factor of ten (or even more) depending on the transition and/or the temperature.  Similarly, values of $\Omega$ need to be calculated over a wide energy range (above thresholds)
in order to obtain convergence of the integral in Eq. (7), as demonstrated in Fig. 7 of Aggarwal and Keenan \cite{ni11a}. 

To delineate resonances, we have performed our calculations of $\Omega$ in the threshold region at over 35000 energies, depending on the ion. Close to thresholds ($\sim$ 0.1 Ryd
above a threshold) the energy mesh is 0.001 Ryd, and away from thresholds is 0.002 Ryd. Thus care has been taken to include as many resonances as possible, and with as fine a
resolution as is computationally feasible. The density and importance of resonances can be appreciated from Figs. 6--8, where we show $\Omega$ in the thresholds region for three
transitions of Ar XVI, namely 1--2 (2s $^2$S$_{1/2}$  -- 2p $^2$P$^o_{1/2}$), 1--3 (2s $^2$S$_{1/2}$ -- 2p $^2$P$^o_{3/2}$),  and 2--3 (2p $^2$P$^o_{1/2}$ -- 2p
$^2$P$^o_{3/2}$). Similar resonances are observed for transitions in other Li-like ions. These resonances affect the values of $\Upsilon$ particularly towards the lower end of
the temperature range.

Our calculated values of $\Upsilon$ are listed in Tables 4 (a--h) over a wide temperature range up to 10$^{7.4}$ K, suitable for applications in solar and other plasmas. As
stated in section 1, there are no similar results available for comparison purposes for most of the ions. Therefore, as for collision strengths we have also calculated values of
$\Upsilon$ from our non-resonant $\Omega$ data from the {\sc fac} code, which are included in Tables 4 (a--h) at the lowest and the highest calculated temperatures  for each
ion. This provides a ready comparison between the two independent calculations, and also gives an estimate of the importance of resonances. For example, at T$_e$ = 10$^{5.5}$ K,
our resonances-resolved values of $\Upsilon$ differ by over 20\% for 120 ($\sim$ 40\%) transitions of Mg X, and in some instances by up to an order of magnitude, see for
example: 2--4, 3--4, and 7--9.  Many of these are resonance transitions, such as 1--5/6/7/8/10/11, and our results from {\sc darc} are invariably  higher due to the inclusion of
resonances. However, at T$_e$ = 10$^{6.9}$ K, only 24 ($\sim$ 9\%) transitions differ by over 20\%, with  a maximum discrepancy of a factor of two for transitions such as:
2--17, 5--17, and 6--18. For the 12--22 (4d $^2$D$_{3/2}$ -- 5f $^2$F$^o_{7/2}$) transition, values of $\Upsilon$ from the {\sc fac} code are higher by a factor of 2.7 because
of a sudden rise in the corresponding values of $\Omega$, as shown for the 10--13 transition in Fig. 5. Overall we may state that the contribution of resonances is more dominant
and appreciable at lower temperatures.  Similar conclusions apply for transitions in other Li-like ions.

The only Li-like ion considered here for which results for $\Upsilon$ are available in the literature is Ar XVI. Whiteford et al. \cite{ar16b} have reported values of $\Upsilon$
for all transitions among the lowest 24 levels over a wide range of electron temperature. They have adopted the standard $R$-matrix code of Berrington et al. \cite{rm1} and have
included one-body relativistic operators in the calculations, which should be sufficient for a moderately heavy ion such as Ar XVI. Furthermore, they included a comparable large
range of partial waves  with angular momentum $J \le$ 58 and resolved resonances in the thresholds region in a narrow energy mesh of $\sim$ 0.002 Ryd. Therefore, their results
for $\Upsilon$ should be comparable to our calculations from the fully relativistic {\sc darc} code. A comparison between the two independent sets of $\Upsilon$ is indeed highly
satisfactory for a {\em majority} of transitions. For only 10 transitions are the differences over 20\%, and the discrepancies are particularly large for seven, namely 7--8,
12--13, 14--15, 19--20, 21--22, 21--24, and 23--24. All of these transitions are forbidden, there is no (major) discrepancy between the $\Omega$ values from the {\sc darc} and
{\sc fac} codes, and there is no large variation of their $\Omega$ values with increasing energy, i.e. $\Omega_B$ is quite stable. Therefore, the differences in the values of
$\Upsilon$ should not be due to the corresponding differences in the values of $\Omega$, but because of the contribution of resonances. However, all of these transitions belong
to the same $n$ and therefore their $\Delta E_{ij}$ are very small, but resonances in collision strengths are neither too dense nor too high in magnitude. In fact, 7--8 (3d
$^2$D$_{3/2}$ --  3d $^2$D$_{5/2}$) is the only transition for which both the thresholds range and resonances are appreciable, as shown in Fig. 9.  

In Fig. 10 we compare values of $\Upsilon$ from the BPRM calculations of Whiteford et al. \cite{ar16b} with our own from the {\sc darc} code for three transitions, namely  7--8
(3d $^2$D$_{3/2}$ --  3d $^2$D$_{5/2}$), 12--13 (4d $^2$D$_{3/2}$ -- 4d $^2$D$_{5/2}$), and 14--15 (4f $^2$F$^o_{5/2}$ -- 4f $^2$F$^o_{7/2}$). At the lowest common temperature
of 5.12$\times$10$^4$ K, values of $\Upsilon$ from the BPRM calculations are higher, by up to a factor of $\sim$30, for the above three and some other transitions, noted above.
As the electron temperature increases the discrepancy between the two calculations decreases significantly, and towards the higher end of the temperature range the two
independent calculations become comparable. The resonances for the 7--8 transition are neither too high in magnitude nor dense (see Fig. 9), and hence cannot make a large
contribution to the determination of $\Upsilon$ as shown by our calculations from {\sc darc}. Clearly, the BPRM calculations of $\Upsilon$ have overestimated the contribution of
resonances for the above noted seven transitions. The main reason for such a large discrepancy between the two $R$-matrix calculations for a few transitions, particularly those
between the degenerate levels of a state, is the limitation of the BPRM method, as recently discussed and demonstrated by Bautista et al. \cite{feiii}. The $jj$ coupling method
adopted in the {\sc darc} code includes fine-structure in the definition of channel coupling and hence gives a better representation of resonances, particularly for transitions
within the  degenerate levels of a state. These near-threshold resonances affect the subsequent results of $\Upsilon$, especially at lower temperatures as shown in Fig. 10. We
discuss this further below.

Another source of higher values of $\Upsilon$ from the BPRM calculations of Whiteford et al. \cite{ar16b}, for some transitions and at lower temperatures, could be  their
inclusion of higher levels of the 1s$n{\ell}{n'}{\ell}'$ configurations. These additional levels in the 226-270 Ryd energy range, well above the highest threshold of our
calculations, i.e. $\sim$ 57 Ryd, may give rise to some resonances which may affect the values of $\Upsilon$. However, it may be true for the calculations of $\Upsilon$ at very
high temperatures, such as T$_e$ = 10$^{7.4}$ K (equivalent to $\sim$ 163 Ryd), but resonances will {\em not} contribute to $\Upsilon$ values at lower temperatures, such as
T$_e$ = 10$^{4.7}$ K (equivalent to $\sim$ 0.32 Ryd) until there are very high spurious resonances. Therefore, we are confident of our results listed in Tables 4 (a--h) which
are assessed to be accurate to better than $\sim$20\% for a majority of transitions and over a wide range of temperature. However, the contribution of higher neglected
resonances, from the 1s$^2$$n{\ell}$ ($n \ge$ 6) and  1s$n{\ell}{n'}{\ell}'$ configurations, may affect these results towards the higher end of the temperature range, as
discussed above, and particularly for transitions involving the levels of the $n$ = 5 configurations.

\begin{flushleft}
{\bf 7. Conclusions}
\end{flushleft}

In this paper we have presented results for energy levels and  radiative rates for four types of transitions (E1, E2, M1, and M2) among the lowest 24 levels of Li-like ions with
12 $\le$ Z $\le$ 20 belonging to the $n \le$ 5 configurations. Additionally, lifetimes of all the levels have been reported, although  measurements or other theoretical results
are not available for comparison for most of the ions/levels.  However, based on a variety of comparisons among various calculations from the {\sc grasp} and {\sc fac} codes,
our energy levels are assessed to be accurate to better than 1\%, and the results for radiative rates, oscillator strengths, line strengths, and lifetimes are assessed to be
accurate to better than 20\% for a majority of strong transitions (levels). Similarly, the accuracy of our results for collision strengths and effective collision strengths is
estimated to be better than 20\% for a majority of the transitions. This accuracy estimate is based on the comparison made between two independent calculations performed with
the {\sc darc} and {\sc fac} codes, as well as with other results available in the literature, particularly for Ar XVI. Additionally, we have considered a large range of partial
waves to achieve convergence of $\Omega$ at all energies, included a wide energy range to accurately calculate the values of $\Upsilon$ up to T$_e$ = 10$^{7.4}$ K, and resolved
resonances in a fine energy mesh to account for their contributions. Hence we see no obvious deficiency in our calculated results. However, the present results for effective
collision strengths for transitions involving the levels of the $n$ = 5 configurations may (perhaps) be improved by the inclusion of the levels of the $n$ = 6 configurations, as
has been noted for the A-values. Finally, we would like to note that inclusion of radiation and Auger damping may reduce the contribution of resonances, particularly at lower
temperatures, for some of the transitions. However, it is certainly true for transitions involving the inner-shell, as demonstrated by Whiteford et al. \cite{ar16b}, but there
is no evidence of similar (significant) effect for the transitions considered in this paper. We believe the present set of complete results for radiative and excitation rates
for eight Li-like ions are probably the best available todate and will be highly useful for the modelling of a variety of plasmas.

\section*{Acknowledgment}

FPK is grateful to  AWE Aldermaston for the award of a William Penney Fellowship and we thank Dr. P.H. Norrington for providing his revised {\sc grasp} and {\sc darc} codes
prior to publication.



\newpage
\begin{flushleft}

\end{flushleft}


\newpage
\begin{flushleft}
{\bf Captions for Figures} \\  \vspace{0.2 cm}
\end{flushleft}

\begin{flushleft}
\begin{tabular}{ll}
Figure 1    &  Partial collision strengths for the 2s $^2$S$_{1/2}$ -- 2p $^2$P$^o_{3/2}$ (1--3) transition of Ar XVI, at five energies of: \\
            & 100 Ryd (circles), 200 Ryd (triangles), 300 Ryd (stars), 400 Ryd (squares), and 500 Ryd (diamonds). \\
Figure 2    & Partial collision strengths for the 2s $^2$S$_{1/2}$ -- 3d $^2$D$_{5/2}$ (1--8) transition of Ar XVI, at five energies of: \\
            & 100 Ryd (circles), 200 Ryd (triangles), 300 Ryd (stars), 400 Ryd (squares), and 500 Ryd (diamonds). \\
Figure 3    & Partial collision strengths for the 2s $^2$S$_{1/2}$ -- 4p $^2$P$^o_{1/2}$ (1--10) transition of Ar XVI, at five energies of: \\
            & 100 Ryd (circles), 200 Ryd (triangles), 300 Ryd (stars), 400 Ryd (squares), and 500 Ryd (diamonds). \\
Figure 4    & Comparison of collision strengths from our calculations from {\sc darc} (continuous curves) \\
            & and {\sc fac} (broken curves) for the 4--6 (circles: 3s $^2$S$_{1/2}$ -- 3p $^2$P$^o_{3/2}$), 6--8 (triangles: 3p $^2$P$^o_{3/2}$ -- 3d $^2$D$_{5/2}$), and  \\
            & 10--12 (stars: 4p $^2$P$^o_{1/2}$ --  4d $^2$D$_{3/2}$) allowed transitions of Ar XVI.\\
Figure 5    & Comparison of collision strengths from our calculations from {\sc darc} (continuous curves) and \\
            & {\sc fac} (broken curves) for the 6--11 (triangles: 3p $^2$P$^o_{3/2}$ -- 4p $^2$P$^o_{3/2}$), 8--13 (stars: 3d $^2$D$_{5/2}$ -- 4d $^2$D$_{5/2}$), and  \\
            & 10--13 (circles: 4p $^2$P$^o_{1/2}$ -- 4d $^2$D$_{5/2}$), forbidden transitions of Ar XVI. \\
Figure 6    & Collision strengths for the 2s $^2$S$_{1/2}$ -- 2p $^2$P$^o_{1/2}$   (1--2) transition of Ar XVI. \\
Figure 7    & Collision strengths for the 2s $^2$S$_{1/2}$ -- 2p $^2$P$^o_{3/2}$  (1--3) transition of Ar XVI. \\
Figure 8    & Collision strengths for the 2p $^2$P$^o_{1/2}$ -- 2p $^2$P$^o_{3/2}$ (2--3) transition of Ar XVI. \\
Figure 9    & Collision strengths for the 3d $^2$D$_{3/2}$ -- 3d $^2$D$_{5/2}$ (7--8) transition of Ar XVI. \\
Figure 10   & Comparison of effective collision strengths for the 7--8 (circles: 3d $^2$D$_{3/2}$ --  3d $^2$D$_{5/2}$), \\
            & 12--13 (triangles: 4d $^2$D$_{3/2}$ -- 4d $^2$D$_{5/2}$), and 14--15 (stars: 4f $^2$F$^o_{5/2}$ -- 4f $^2$F$^o_{7/2}$)\\
            & transitions of Ar XVI. Continuous and broken curves are from the present {\sc darc} and \\
	    & earlier BPRM codes \cite{ar16b}, respectively. \\

\end{tabular}
\end{flushleft}


\newpage
\clearpage
\begin{figure*}
\includegraphics[angle=-90,width=0.90\textwidth]{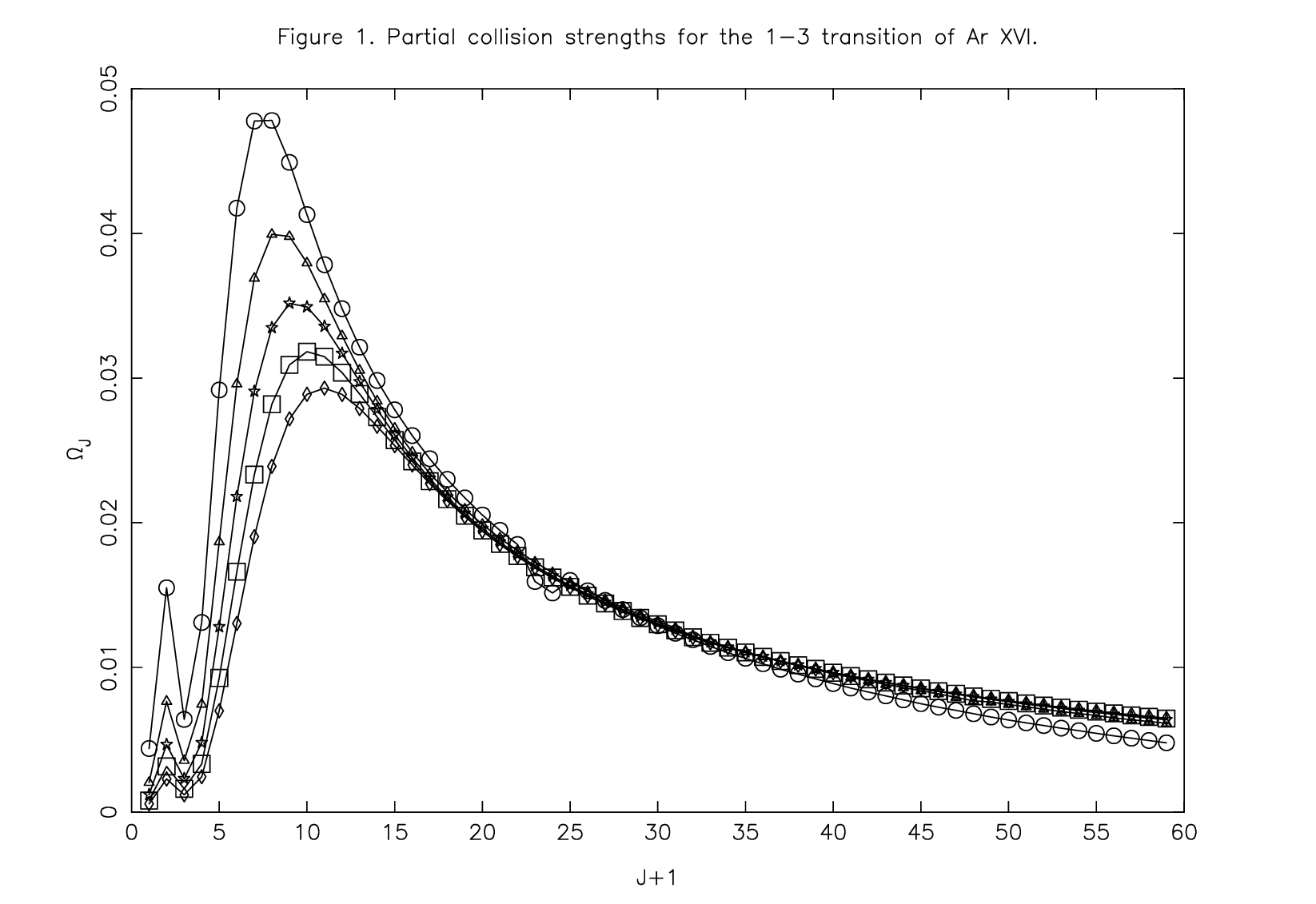}
\caption{Partial collision strengths for the 2s $^2$S$_{1/2}$ -- 2p $^2$P$^o_{3/2}$ (1--3) transition of Ar XVI,
at five energies of: 100 Ryd (circles), 200 Ryd (triangles), 300 Ryd (stars), 400 Ryd (squares), and 500 Ryd (diamonds).}
\end{figure*}


\setcounter{figure} {1}
\begin{figure*}
\includegraphics[angle=-90,width=0.90\textwidth]{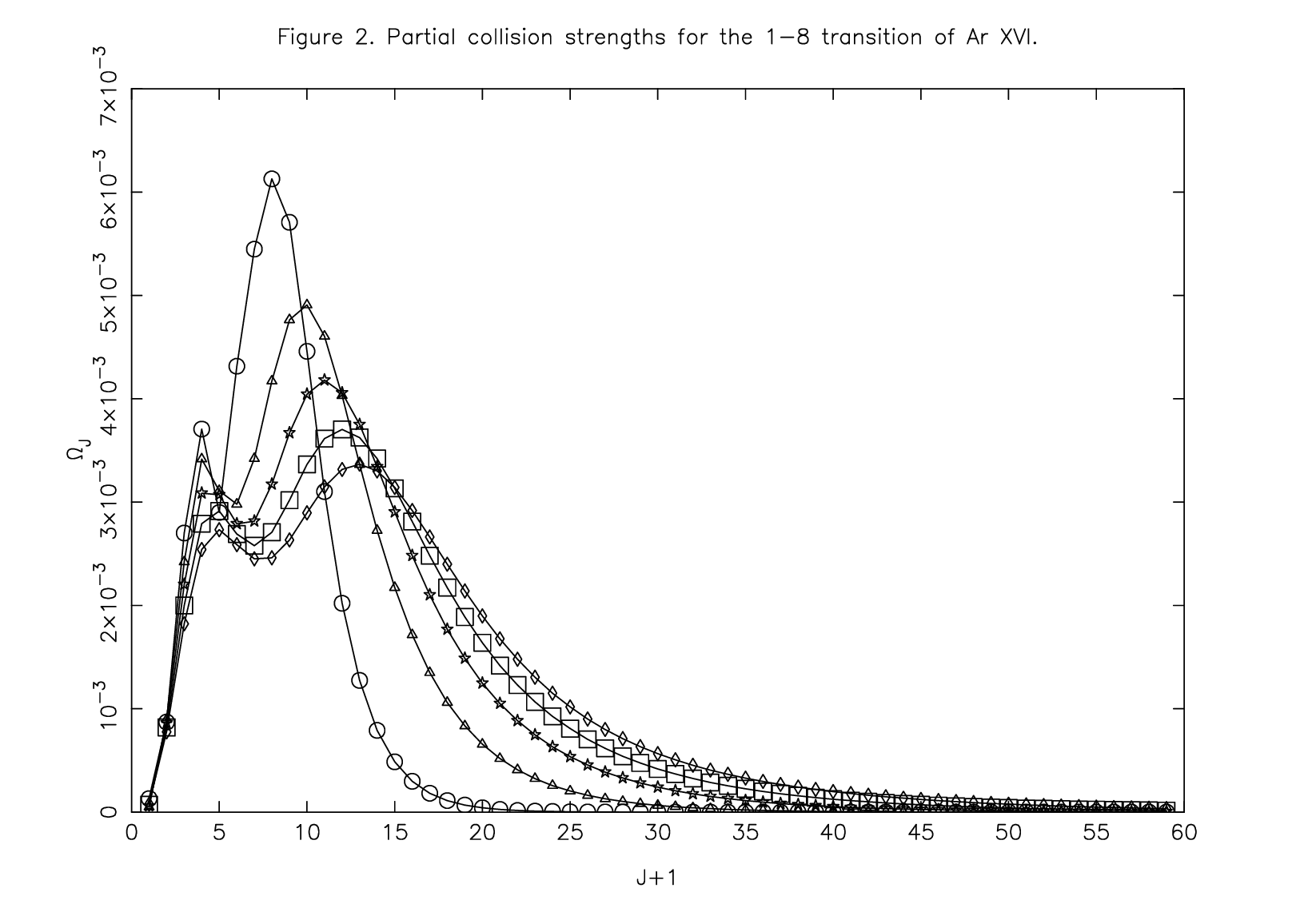}
\caption{Partial collision strengths for the 2s $^2$S$_{1/2}$ -- 3d $^2$D$_{5/2}$ (1--8) transition of Ar XVI,
at five energies of: 100 Ryd (circles), 200 Ryd (triangles), 300 Ryd (stars), 400 Ryd (squares), and 500 Ryd (diamonds).}
\end{figure*}


\clearpage
\setcounter{figure} {2}
\begin{figure*}
\includegraphics[angle=-90,width=0.90\textwidth]{fig3.ps}
\caption{Partial collision strengths for the 2s $^2$S$_{1/2}$ -- 4p $^2$P$^o_{1/2}$ (1--10) transition of Ar XVI,
at five energies of: 100 Ryd (circles), 200 Ryd (triangles), 300 Ryd (stars), 400 Ryd (squares), and 500 Ryd (diamonds).}
\end{figure*}


\setcounter{figure} {3}
\begin{figure*}
\includegraphics[angle=-90,width=0.90\textwidth]{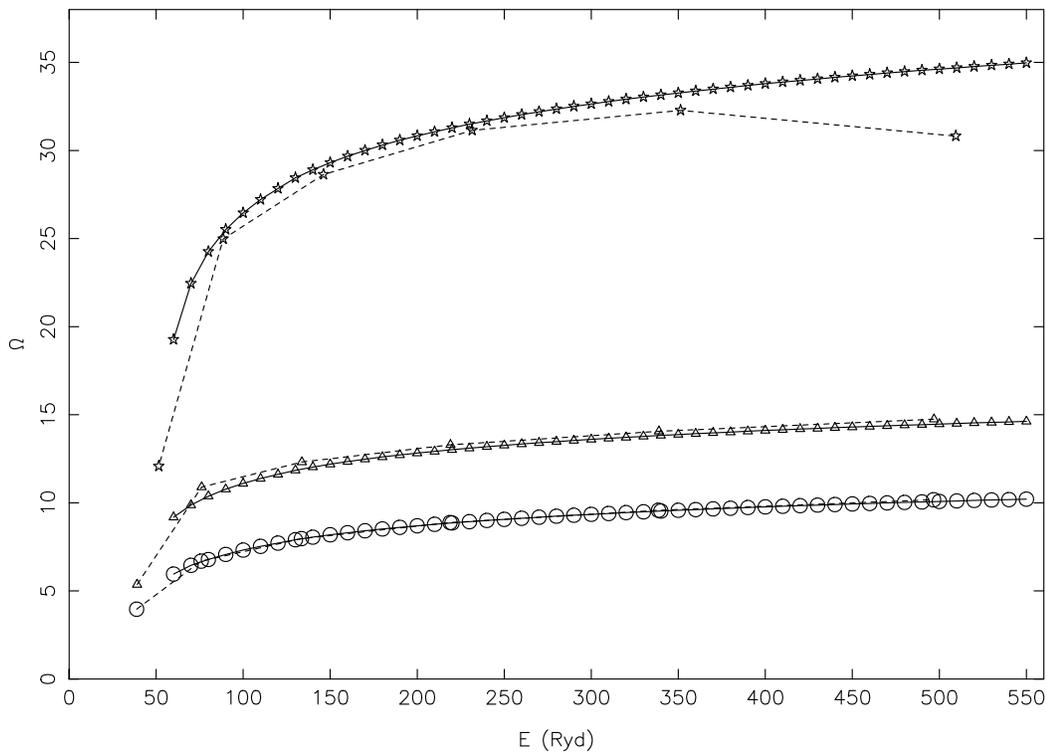}
\caption{Comparison of collision strengths from our calculations from {\sc darc} (continuous curves) and {\sc fac} (broken curves) for the
  4--6 (circles: 3s $^2$S$_{1/2}$ -- 3p $^2$P$^o_{3/2}$), 6--8 (triangles: 3p $^2$P$^o_{3/2}$ -- 3d $^2$D$_{5/2}$), and 10--12 (stars: 4p $^2$P$^o_{1/2}$ --  4d $^2$D$_{3/2}$) 
  allowed transitions of Ar XVI.}
\end{figure*}


\setcounter{figure} {4}
\begin{figure*}
\includegraphics[angle=-90,width=0.90\textwidth]{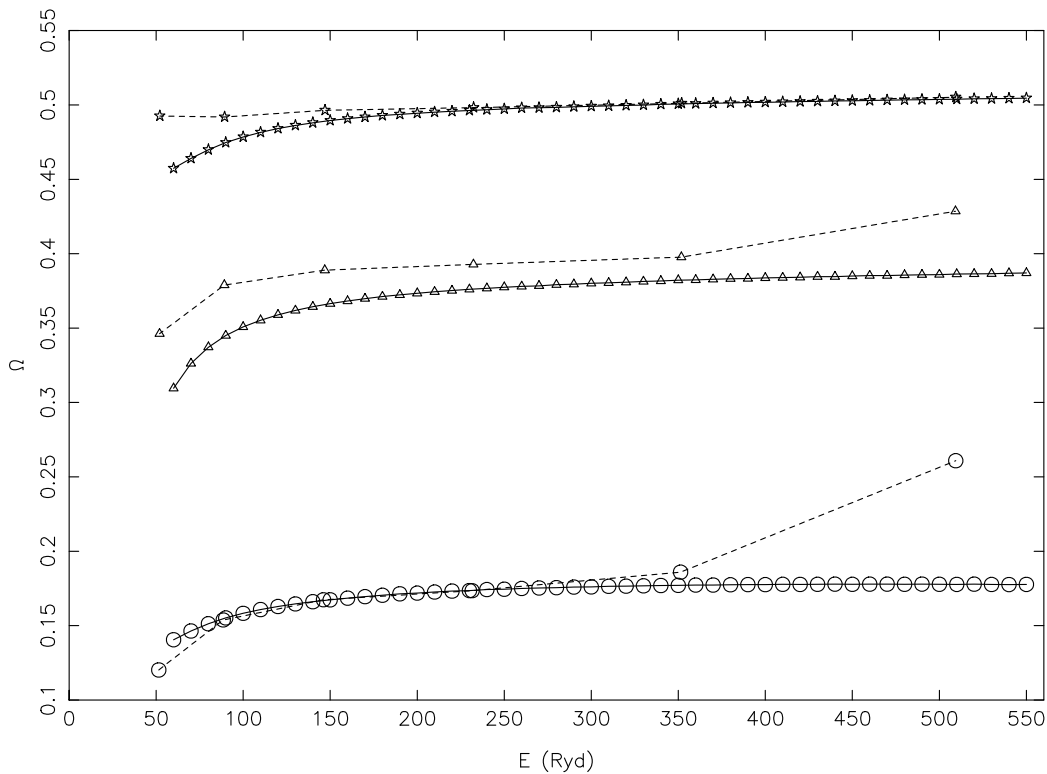}
\caption{Comparison of collision strengths from our calculations from {\sc darc} (continuous curves) and {\sc fac} (broken curves) for the
 6--11 (triangles: 3p $^2$P$^o_{3/2}$ -- 4p $^2$P$^o_{3/2}$), 8--13 (stars: 3d $^2$D$_{5/2}$ -- 4d $^2$D$_{5/2}$), and 
10--13 (circles: 4p $^2$P$^o_{1/2}$ -- 4d $^2$D$_{5/2}$), forbidden transitions of Ar XVI.}
\end{figure*}


\setcounter{figure} {5}
\begin{figure*}
\includegraphics[angle=+90,width=0.90\textwidth]{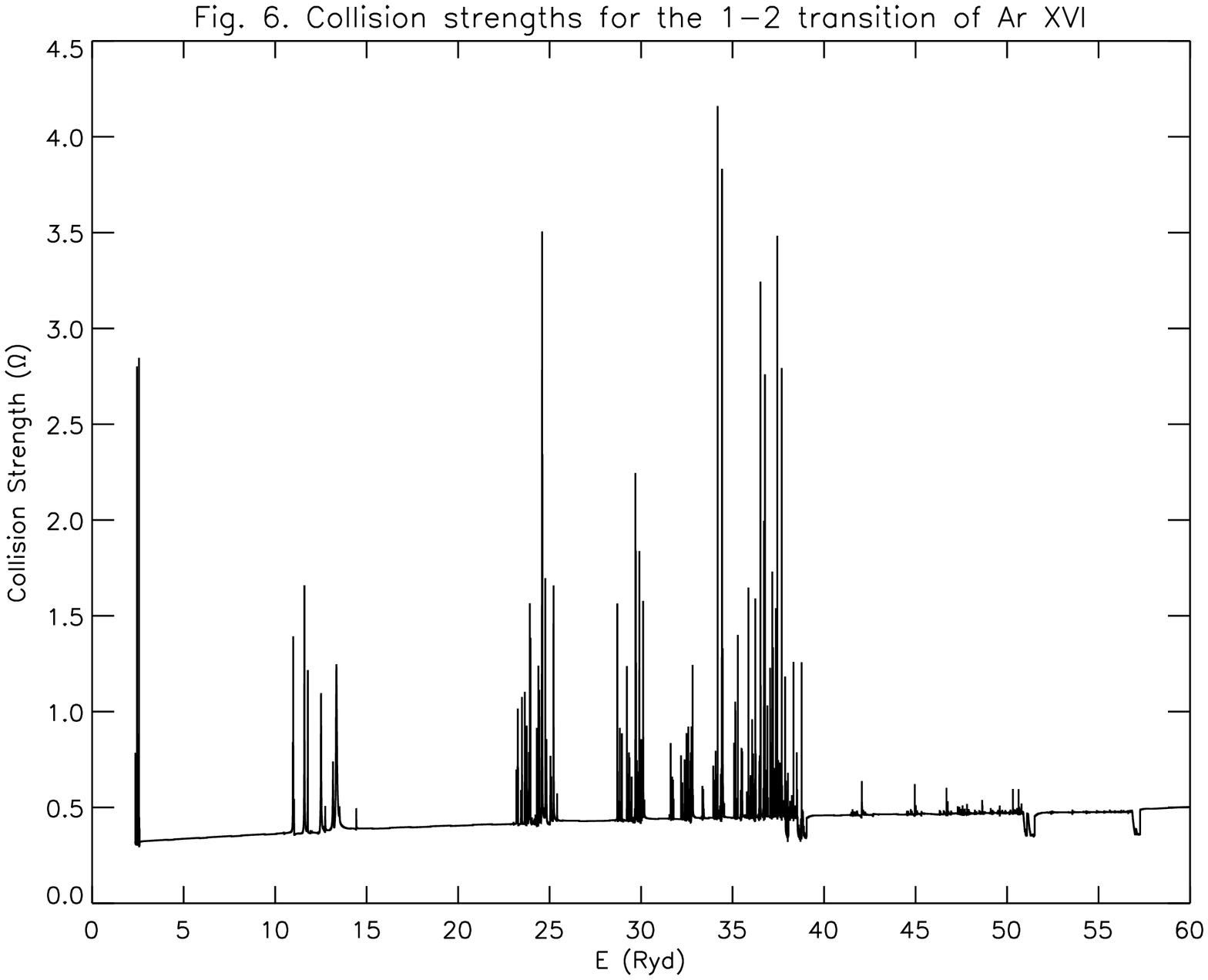}
\caption{Collision strengths for the 2s $^2$S$_{1/2}$ -- 2p $^2$P$^o_{1/2}$   (1--2) transition of Ar XVI.}
\end{figure*}


\setcounter{figure} {6}
\begin{figure*}
\includegraphics[angle=+90,width=0.90\textwidth]{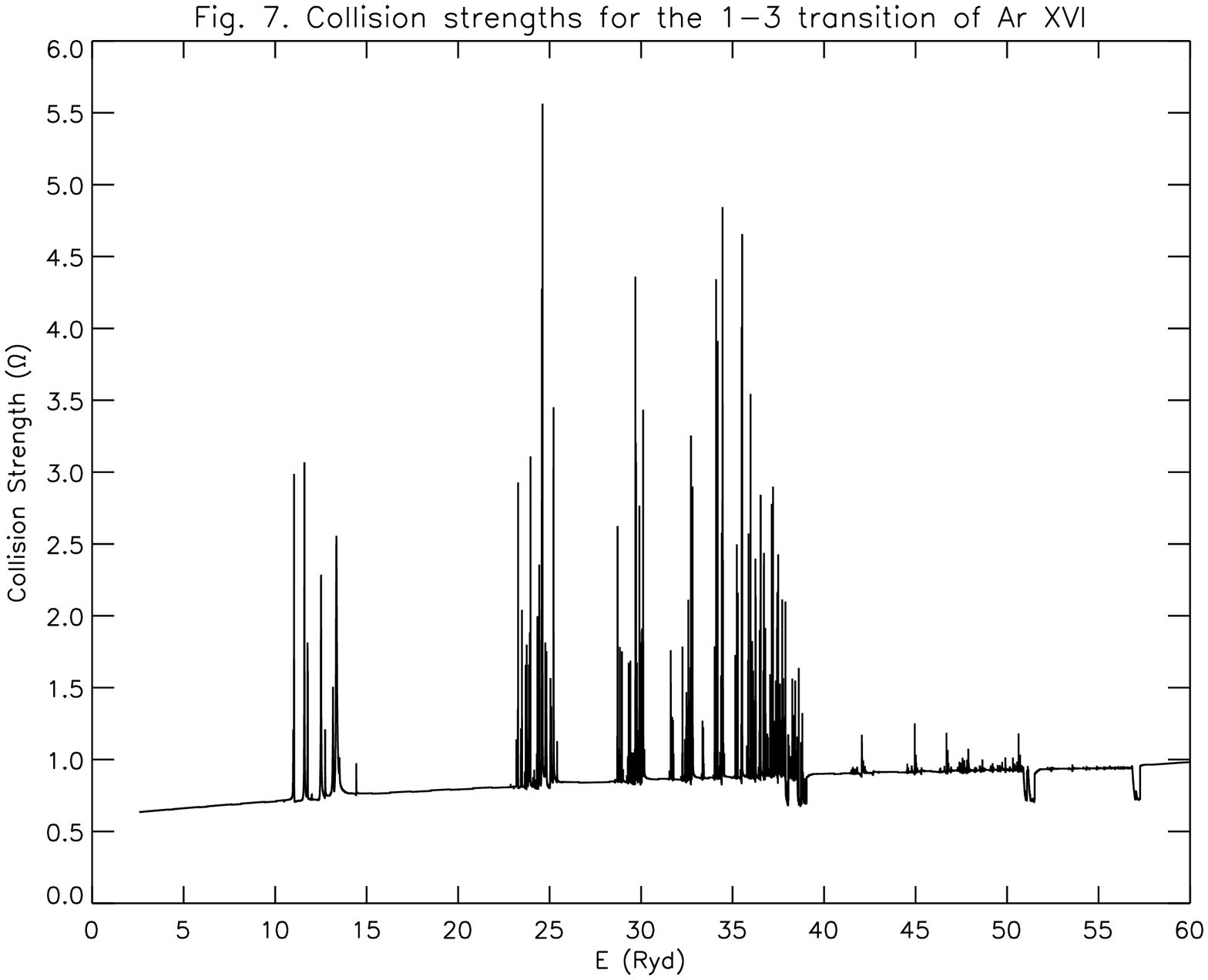}
\caption{Collision strengths for the 2s $^2$S$_{1/2}$ -- 2p $^2$P$^o_{3/2}$  (1--3) transition of Ar XVI.}
\end{figure*}


\setcounter{figure} {7}
\begin{figure*}
\includegraphics[angle=+90,width=0.90\textwidth]{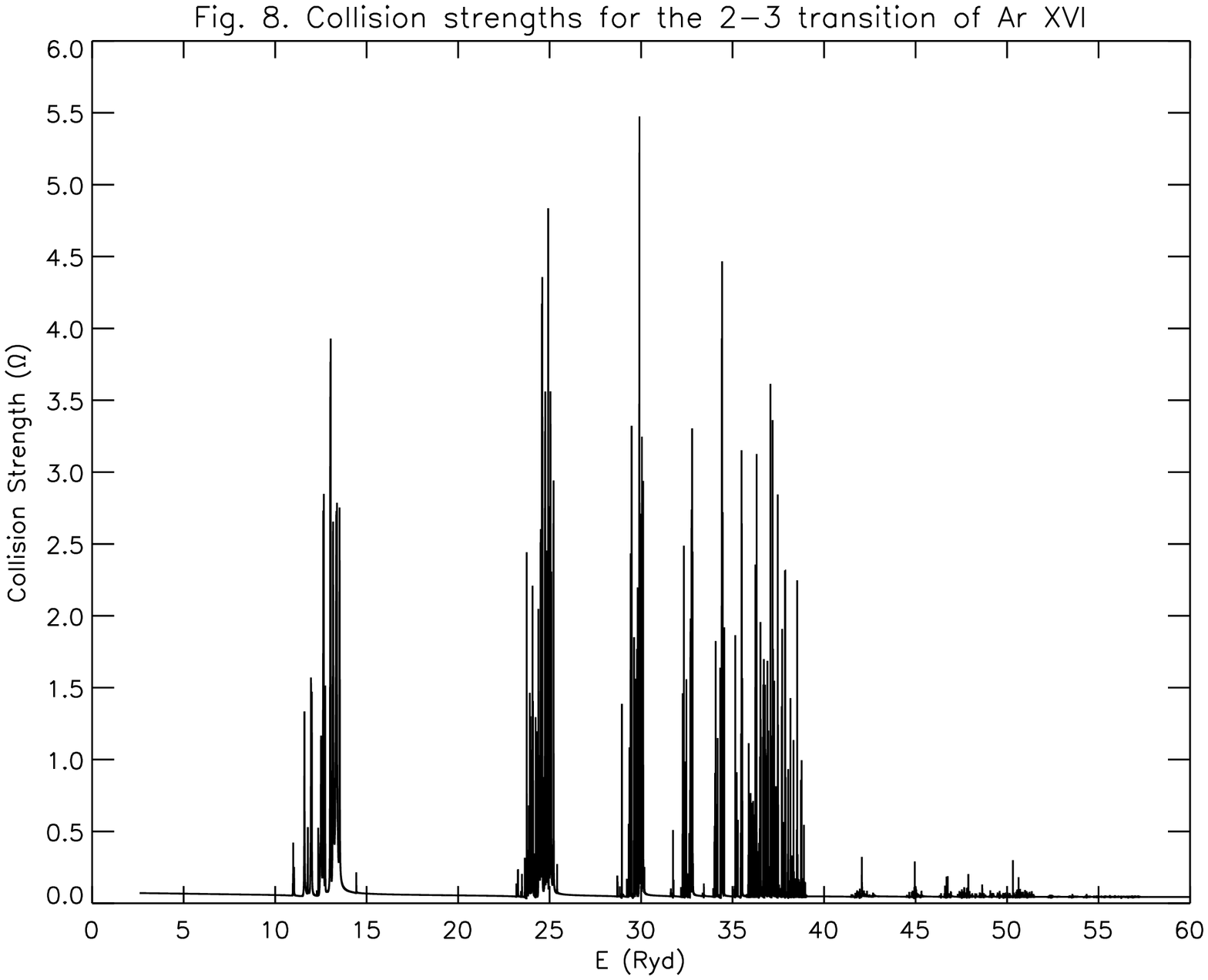}
\caption{Collision strengths for the 2p $^2$P$^o_{1/2}$ -- 2p $^2$P$^o_{3/2}$ (2--3) transition of Ar XVI.}
\end{figure*}

\setcounter{figure} {8}
\begin{figure*}
\includegraphics[angle=+90,width=0.90\textwidth]{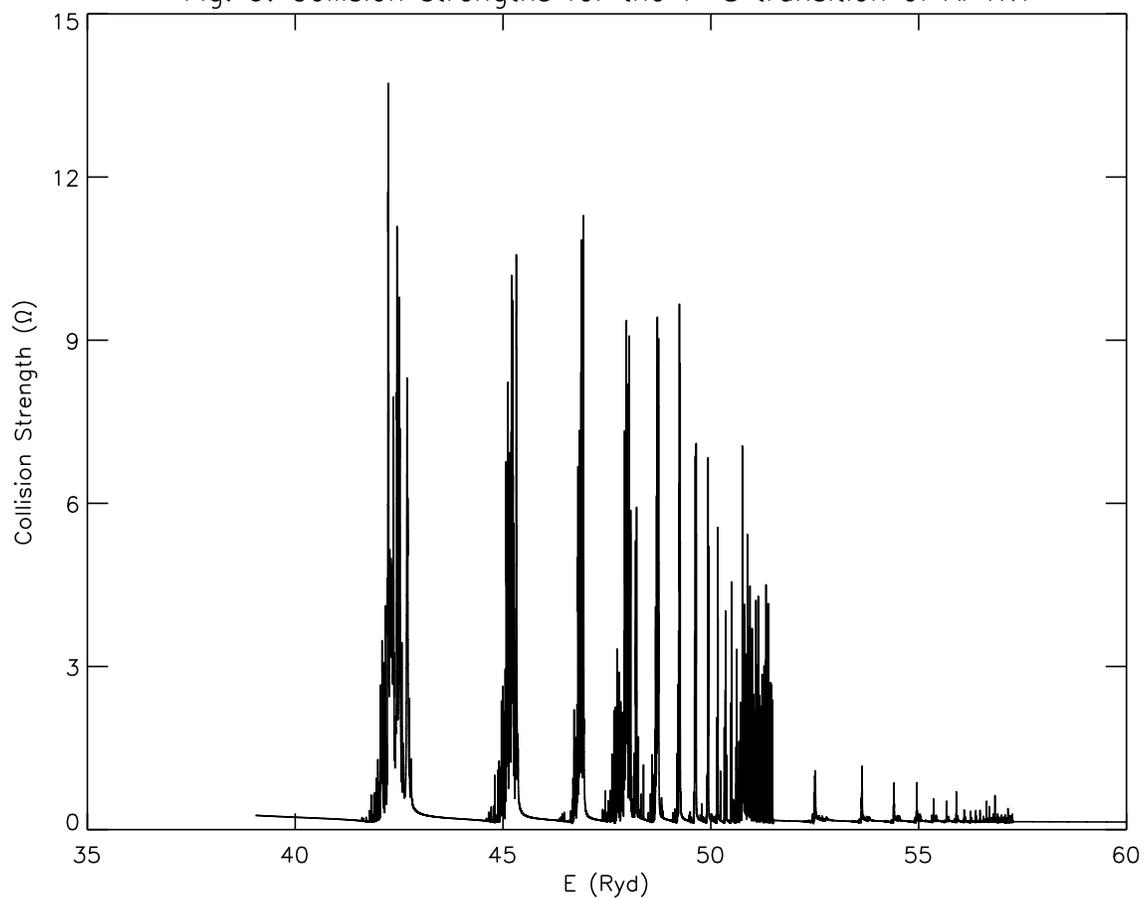}
\caption{Collision strengths for the 3d $^2$D$_{3/2}$ -- 3d $^2$D$_{5/2}$ (7--8) transition of Ar XVI.}
\end{figure*}

\setcounter{figure} {9}
\begin{figure*}
\includegraphics[angle=-90,width=0.90\textwidth]{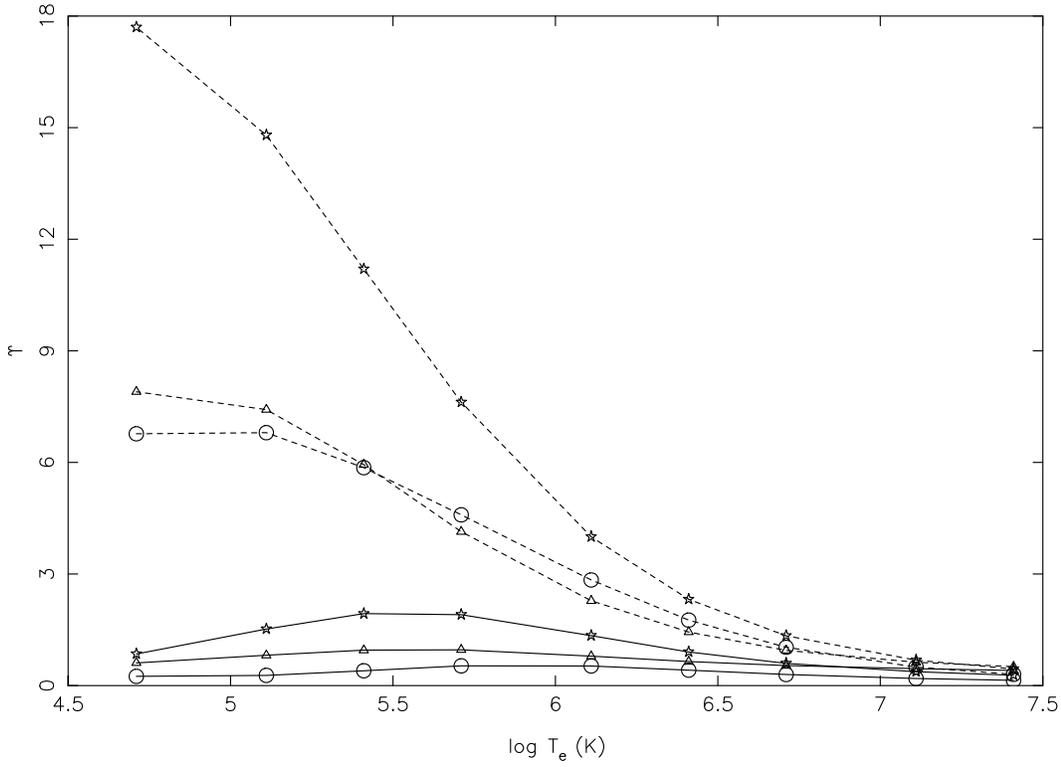}
\caption{Comparison of effective collision strengths for the 7--8 (circles: 3d $^2$D$_{3/2}$ --  3d $^2$D$_{5/2}$), 
12--13 (triangles: 4d $^2$D$_{3/2}$ -- 4d $^2$D$_{5/2}$), and 14--15 (stars: 4f $^2$F$^o_{5/2}$ -- 4f $^2$F$^o_{7/2}$)
 transitions of Ar XVI. Continuous and broken curves are from the present {\sc darc} and earlier BPRM codes \cite{ar16b}, respectively.}
\end{figure*}


\newpage
\clearpage

\begin{table*}
\begin{flushleft}
{\bf Table A.} Comparison of radiative rates (A- values, s$^{-1}$) for some transitions of Mg X.  ($a{\pm}b \equiv a{\times}$10$^{{\pm}b}$). \\
\end{flushleft}
\begin{tabular}{rrrrrr}

\hline

\hline

$i$ & $j$ & A (GRASP) & A (FAC) & A (BPRM) & f (GRASP)        \\

\hline
    1 &    2 &  7.1756$+$08 &  7.2480$+$08 &  6.8880$+$08 &  4.1444$-$02  \\
    1 &    3 &  7.7364$+$08 &  7.8020$+$08 &  7.5850$+$08 &  8.5140$-$02  \\
    1 &    5 &  2.1364$+$11 &  2.2050$+$11 &  2.1740$+$11 &  1.0751$-$01  \\
    1 &    6 &  2.1220$+$11 &  2.1910$+$11 &  2.1560$+$11 &  2.1328$-$01  \\
    1 &   10 &  9.7003$+$10 &  1.0610$+$11 &  9.5610$+$10 &  2.8233$-$02  \\
    1 &   11 &  9.6513$+$10 &  1.0570$+$11 &  9.5050$+$10 &  5.6157$-$02  \\
    1 &   17 &  5.0573$+$10 &  6.3780$+$10 &  4.9230$+$10 &  1.1938$-$02  \\
    1 &   18 &  5.0340$+$10 &  6.3520$+$10 &  4.8990$+$10 &  2.3762$-$02  \\
    2 &    7 &  5.5804$+$11 &  5.6220$+$11 &  5.5730$+$11 &  6.6908$-$01  \\
    2 &   12 &  1.8354$+$11 &  1.8970$+$11 &  1.8230$+$11 &  1.2299$-$01  \\
    2 &   19 &  8.4912$+$10 &  9.6940$+$10 &  8.4130$+$10 &  4.5616$-$02  \\
    3 &    8 &  6.6798$+$11 &  6.7310$+$11 &  6.6660$+$11 &  6.0341$-$01  \\
    3 &   13 &  2.1942$+$11 &  2.2670$+$11 &  2.1770$+$11 &  1.1067$-$01  \\
    3 &   20 &  1.0141$+$11 &  1.1560$+$11 &  1.0040$+$11 &  4.0995$-$02  \\
    4 &    5 &  9.0225$+$07 &  9.0380$+$07 &  8.4280$+$07 &  6.9443$-$02  \\
    4 &    6 &  9.7710$+$07 &  9.7690$+$07 &  9.3350$+$07 &  1.4277$-$01  \\
    4 &   10 &  2.6720$+$10 &  2.7350$+$10 &  2.6570$+$10 &  1.1598$-$01  \\
    4 &   11 &  2.6506$+$10 &  2.7140$+$10 &  2.6310$+$10 &  2.2972$-$01  \\
    4 &   17 &  1.5113$+$10 &  1.7210$+$10 &  1.4940$+$10 &  3.2247$-$02  \\
    4 &   18 &  1.5016$+$10 &  1.7100$+$10 &  1.4830$+$10 &  6.4043$-$02  \\
    5 &    7 &  3.0955$+$06 &		   &  3.2370$+$06 &  3.4071$-$02  \\
    5 &   12 &  5.7952$+$10 &  5.8890$+$10 &  5.7960$+$10 &  5.7352$-$01  \\
    5 &   19 &  2.8734$+$10 &  3.0740$+$10 &  2.8740$+$10 &  1.3543$-$01  \\
    6 &    8 &  3.1978$+$06 &		   &  3.2100$+$06 &  2.9176$-$02  \\
    6 &   13 &  6.9546$+$10 &  7.0660$+$10 &  6.9570$+$10 &  5.1810$-$01  \\
    6 &   20 &  3.4430$+$10 &  3.6800$+$10 &  3.4440$+$10 &  1.2204$-$01  \\
    7 &   11 &  4.2619$+$08 &  4.3130$+$08 &  1.1110$+$11 &  2.2959$-$03  \\
    7 &   14 &  1.2895$+$11 &  1.2840$+$11 &  1.2920$+$11 &  1.0155$+$00  \\
    7 &   18 &  1.8239$+$08 &  2.0810$+$08 &  4.9120$+$05 &  4.5193$-$04  \\
    7 &   21 &  4.2561$+$10 &  4.2570$+$10 &  4.2640$+$10 &  1.5678$-$01  \\
    8 &   14 &  9.2022$+$09 &  9.1640$+$09 &  9.2210$+$09 &  4.8379$-$02  \\
    8 &   15 &  1.3806$+$11 &  1.3750$+$11 &  1.3840$+$11 &  9.6752$-$01  \\
    8 &   21 &  3.0346$+$09 &  3.0340$+$09 &  3.0390$+$09 &  7.4592$-$03  \\
    8 &   22 &  4.5553$+$10 &  4.5580$+$10 &  4.5620$+$10 &  1.4928$-$01  \\
    9 &   10 &  2.0896$+$07 &  1.9240$+$07 &  2.6330$+$07 &  9.5744$-$02  \\
    9 &   11 &  2.2665$+$07 &  2.0860$+$07 &  2.8960$+$07 &  1.9688$-$01  \\
    9 &   17 &  6.1436$+$09 &  6.4780$+$09 &  6.0070$+$09 &  1.2764$-$01  \\
    9 &   18 &  6.0898$+$09 &  6.4230$+$09 &  5.9410$+$09 &  2.5258$-$01  \\
   10 &   12 &  9.5421$+$05 &		   &  1.1860$+$06 &  6.0440$-$02  \\
   10 &   19 &  1.1941$+$10 &  1.2560$+$10 &  1.1890$+$10 &  5.5482$-$01  \\
   11 &   13 &  9.8431$+$05 &		   &  1.1860$+$06 &  5.1730$-$02  \\
   11 &   20 &  1.4349$+$10 &  1.5080$+$10 &  1.4300$+$10 &  5.0164$-$01  \\
   12 &   14 &  1.7043$+$02 &		   &  2.5830$+$02 &  2.0074$-$03  \\
   12 &   18 &  2.2549$+$08 &  2.2940$+$08 &  1.1610$+$10 &  5.6826$-$03  \\
   12 &   21 &  2.4136$+$10 &  2.3840$+$10 &  2.4120$+$10 &  8.8681$-$01  \\
   13 &   14 &  3.3940$+$00 &		   &  4.6590$+$00 &  6.2432$-$05  \\
   13 &   15 &  1.0349$+$02 &		   &  1.4940$+$02 &  1.5822$-$03  \\
   13 &   21 &  1.7239$+$09 &  1.7030$+$09 &  1.7230$+$09 &  4.2279$-$02  \\
   13 &   22 &  2.5853$+$10 &  2.5540$+$10 &  2.5840$+$10 &  8.4514$-$01  \\
   14 &   23 &  4.1057$+$10 &  4.1040$+$10 &  4.1040$+$10 &  1.3452$+$00  \\
   16 &   17 &  6.7463$+$06 &		   &  9.5450$+$06 &  1.2141$-$01  \\
   16 &   18 &  7.3233$+$06 &		   &  1.0470$+$07 &  2.4969$-$01  \\
   17 &   19 &  3.4348$+$05 &		   &  4.5290$+$05 &  8.3996$-$02  \\
   18 &   20 &  3.5413$+$05 &		   &  4.5410$+$05 &  7.1879$-$02  \\
   19 &   21 &  9.9345$+$01 &		   &  1.4580$+$02 &  3.9179$-$03  \\
   20 &   21 &  2.1863$+$00 &		   &  2.8820$+$00 &  1.2598$-$04  \\
   20 &   22 &  6.2785$+$01 &		   &  8.7420$+$01 &  3.1294$-$03  \\
   21 &   23 &  3.1503$-$01 &		   &  5.4940$-$01 &  3.6495$-$04  \\
\hline   
\end{tabular}

\begin{flushleft}
{\small
GRASP: Present results from the {\sc grasp} code \\
FAC:	Present results from the {\sc fac} code \\
BPRM:  Nahar \cite{sn} \\
}
\end{flushleft}
\end{table*}


\newpage
\clearpage

\begin{table*}
\begin{flushleft}
{\bf Table B.} Comparison of radiative rates (A- values, s$^{-1}$) for some transitions of Al XI.  ($a{\pm}b \equiv a{\times}$10$^{{\pm}b}$). \\
\end{flushleft}
\begin{tabular}{rrrrrr}

\hline

\hline

$i$ & $j$ & A (GRASP) & A (FAC) & A (BPRM) & f (GRASP)        \\

\hline
    1 &    2 &  7.9134$+$08 &  7.9850$+$08 &  7.6000$+$08 &  3.7806$-$02  \\
    1 &    3 &  8.7385$+$08 &  8.8010$+$08 &  8.6070$+$08 &  7.8319$-$02  \\
    1 &    5 &  3.1534$+$11 &  3.2450$+$11 &  3.2110$+$11 &  1.1052$-$01  \\
    1 &    6 &  3.1282$+$11 &  3.2210$+$11 &  3.1800$+$11 &  2.1892$-$01  \\
    1 &   10 &  1.4250$+$11 &  1.5450$+$11 &  1.3820$+$11 &  2.8763$-$02  \\
    1 &   11 &  1.4166$+$11 &  1.5370$+$11 &  1.3730$+$11 &  5.7157$-$02  \\
    1 &   17 &  7.4182$+$10 &  9.1400$+$10 &  7.1930$+$10 &  1.2127$-$02  \\
    1 &   18 &  7.3788$+$10 &  9.0990$+$10 &  7.1520$+$10 &  2.4120$-$02  \\
    2 &    7 &  8.1567$+$11 &  8.2110$+$11 &  8.1470$+$11 &  6.7043$-$01  \\
    2 &   12 &  2.6782$+$11 &  2.7590$+$11 &  2.6640$+$11 &  1.2291$-$01  \\
    2 &   19 &  1.2382$+$11 &  1.3960$+$11 &  1.2260$+$11 &  4.5544$-$02  \\
    3 &    8 &  9.7585$+$11 &  9.8270$+$11 &  9.7400$+$11 &  6.0488$-$01  \\
    3 &   13 &  3.1991$+$11 &  3.2960$+$11 &  3.1790$+$11 &  1.1059$-$01  \\
    3 &   20 &  1.4775$+$11 &  1.6630$+$11 &  1.4630$+$11 &  4.0920$-$02  \\
    4 &    5 &  1.0001$+$08 &  1.0010$+$08 &  9.3020$+$07 &  6.3408$-$02  \\
    4 &    6 &  1.1107$+$08 &  1.1100$+$08 &  1.0620$+$08 &  1.3150$-$01  \\
    4 &   10 &  3.9741$+$10 &  4.0620$+$10 &  3.9210$+$10 &  1.1962$-$01  \\
    4 &   11 &  3.9366$+$10 &  4.0250$+$10 &  3.8750$+$10 &  2.3650$-$01  \\
    4 &   17 &  2.2365$+$10 &  2.5140$+$10 &  2.2080$+$10 &  3.2953$-$02  \\
    4 &   18 &  2.2197$+$10 &  2.4960$+$10 &  2.1890$+$10 &  6.5365$-$02  \\
    5 &    7 &  3.8299$+$06 &              &  4.0140$+$06 &  3.2197$-$02  \\
    5 &   12 &  8.4940$+$10 &  8.6160$+$10 &  8.4980$+$10 &  5.7592$-$01  \\
    5 &   19 &  4.2050$+$10 &  4.4680$+$10 &  4.2000$+$10 &  1.3565$-$01  \\
    6 &    8 &  3.7927$+$06 &              &  3.7910$+$06 &  2.7186$-$02  \\
    6 &   13 &  1.0193$+$11 &  1.0340$+$11 &  1.0200$+$11 &  5.2067$-$01  \\
    6 &   20 &  5.0371$+$10 &  5.3470$+$10 &  5.0320$+$10 &  1.2228$-$01  \\
    7 &   11 &  6.1532$+$08 &  6.2360$+$08 &  1.6230$+$11 &  2.2595$-$03  \\
    7 &   14 &  1.8881$+$11 &  1.8820$+$11 &  1.8920$+$11 &  1.0154$+$00  \\
    7 &   18 &  2.6345$+$08 &  2.9850$+$08 &  5.6600$+$05 &  4.4551$-$04  \\
    7 &   21 &  6.2321$+$10 &  6.2360$+$10 &  6.2410$+$10 &  1.5677$-$01  \\
    8 &   14 &  1.3472$+$10 &  1.3420$+$10 &  1.3500$+$10 &  4.8376$-$02  \\
    8 &   15 &  2.0213$+$11 &  2.0150$+$11 &  2.0250$+$11 &  9.6746$-$01  \\
    8 &   21 &  4.4417$+$09 &  4.4420$+$09 &  4.4460$+$09 &  7.4573$-$03  \\
    8 &   22 &  6.6688$+$10 &  6.6740$+$10 &  6.6760$+$10 &  1.4927$-$01  \\
    9 &   10 &  2.3209$+$07 &  2.1520$+$07 &  3.5820$+$07 &  8.7454$-$02  \\
    9 &   11 &  2.5831$+$07 &  2.3950$+$07 &  4.0180$+$07 &  1.8142$-$01  \\
    9 &   17 &  9.1736$+$09 &  9.6260$+$09 &  8.8860$+$09 &  1.3191$-$01  \\
    9 &   18 &  9.0790$+$09 &  9.5300$+$09 &  8.7700$+$09 &  2.6051$-$01  \\
   10 &   12 &  1.1794$+$06 &              &  1.4090$+$06 &  5.7104$-$02  \\
   10 &   19 &  1.7527$+$10 &  1.8310$+$10 &  1.7450$+$10 &  5.5776$-$01  \\
   11 &   13 &  1.1656$+$06 &              &  1.3400$+$06 &  4.8183$-$02  \\
   11 &   20 &  2.1066$+$10 &  2.2000$+$10 &  2.0990$+$10 &  5.0478$-$01  \\
   12 &   14 &  2.8264$+$02 &              &  4.0410$+$02 &  2.0924$-$03  \\
   12 &   18 &  3.2608$+$08 &  3.3230$+$08 &  1.7020$+$10 &  5.6006$-$03  \\
   12 &   21 &  3.5339$+$10 &  3.4980$+$10 &  3.5320$+$10 &  8.8667$-$01  \\
   13 &   14 &  4.3011$+$00 &              &  5.3750$+$00 &  5.9492$-$05  \\
   13 &   15 &  1.5479$+$02 &              &  2.0820$+$02 &  1.5934$-$03  \\
   13 &   21 &  2.5241$+$09 &  2.4980$+$09 &  2.5230$+$09 &  4.2283$-$02  \\
   13 &   22 &  3.7852$+$10 &  3.7470$+$10 &  3.7830$+$10 &  8.4512$-$01  \\
   14 &   23 &  6.0114$+$10 &  6.0090$+$10 &  6.0090$+$10 &  1.3451$+$00  \\
   16 &   17 &  7.5020$+$06 &              &  1.1960$+$07 &  1.1092$-$01  \\
   16 &   18 &  8.3580$+$06 &              &  1.3420$+$07 &  2.3013$-$01  \\
   17 &   19 &  4.2440$+$05 &              &  5.7530$+$05 &  7.9357$-$02  \\
   18 &   20 &  4.1913$+$05 &              &  5.5120$+$05 &  6.6944$-$02  \\
   19 &   21 &  1.6187$+$02 &              &  2.2620$+$02 &  4.0598$-$03  \\
   20 &   21 &  2.7649$+$00 &              &  3.3620$+$00 &  1.1997$-$04  \\
   20 &   22 &  9.2622$+$01 &              &  1.2150$+$02 &  3.1371$-$03  \\
   21 &   23 &  7.8832$-$01 &              &  1.1930$+$00 &  4.3633$-$04  \\
\hline   
\end{tabular}

\begin{flushleft}
{\small
GRASP: Present results from the {\sc grasp} code \\
FAC:	Present results from the {\sc fac} code \\
BPRM:  Nahar \cite{sn} \\
}
\end{flushleft}
\end{table*}


\newpage
\clearpage

\begin{table*}
\begin{flushleft}
{\bf Table C.} Comparison of radiative rates (A- values, s$^{-1}$) for some transitions of P XIII.  ($a{\pm}b \equiv a{\times}$10$^{{\pm}b}$). \\
\end{flushleft}
\begin{tabular}{rrrrr}

\hline

\hline

$i$ & $j$ & A (GRASP) & A (FAC) &  f (GRASP)        \\

\hline
    1 &    2 &  9.4063$+$08 &  9.4780$+$08 &  3.2166$-$02  \\
    1 &    3 &  1.1043$+$09 &  1.1100$+$09 &  6.8067$-$02  \\
    1 &    5 &  6.2262$+$11 &  6.3770$+$11 &  1.1527$-$01  \\
    1 &    6 &  6.1595$+$11 &  6.3150$+$11 &  2.2753$-$01  \\
    1 &   10 &  2.7929$+$11 &  2.9890$+$11 &  2.9579$-$02  \\
    1 &   11 &  2.7712$+$11 &  2.9690$+$11 &  5.8656$-$02  \\
    1 &   17 &  1.4505$+$11 &  1.7260$+$11 &  1.2415$-$02  \\
    1 &   18 &  1.4406$+$11 &  1.7160$+$11 &  2.4652$-$02  \\
    2 &    7 &  1.5872$+$12 &  1.5960$+$12 &  6.7234$-$01  \\
    2 &   12 &  5.1975$+$11 &  5.3300$+$11 &  1.2275$-$01  \\
    2 &   19 &  2.4009$+$11 &  2.6560$+$11 &  4.5425$-$02  \\
    3 &    8 &  1.8966$+$12 &  1.9080$+$12 &  6.0718$-$01  \\
    3 &   13 &  6.1976$+$11 &  6.3570$+$11 &  1.1044$-$01  \\
    3 &   20 &  2.8589$+$11 &  3.1570$+$11 &  4.0792$-$02  \\
    4 &    5 &  1.1985$+$08 &  1.2000$+$08 &  5.4039$-$02  \\
    4 &    6 &  1.4200$+$08 &  1.4180$+$08 &  1.1457$-$01  \\
    4 &   10 &  7.9387$+$10 &  8.0940$+$10 &  1.2536$-$01  \\
    4 &   11 &  7.8381$+$10 &  7.9950$+$10 &  2.4685$-$01  \\
    4 &   17 &  4.4330$+$10 &  4.8880$+$10 &  3.4048$-$02  \\
    4 &   18 &  4.3890$+$10 &  4.8410$+$10 &  6.7352$-$02  \\
    5 &   12 &  1.6597$+$11 &  1.6790$+$11 &  5.7941$-$01  \\
    5 &   19 &  8.1979$+$10 &  8.6250$+$10 &  1.3596$-$01  \\
    6 &   13 &  1.9918$+$11 &  2.0150$+$11 &  5.2473$-$01  \\
    6 &   20 &  9.8138$+$10 &  1.0310$+$11 &  1.2263$-$01  \\
    7 &   11 &  1.1732$+$09 &  1.1910$+$09 &  2.2012$-$03  \\
    7 &   14 &  3.6842$+$11 &  3.6760$+$11 &  1.0150$+$00  \\
    7 &   18 &  5.0272$+$08 &  5.6250$+$08 &  4.3520$-$04  \\
    7 &   21 &  1.2161$+$11 &  1.2170$+$11 &  1.5675$-$01  \\
    8 &   14 &  2.6274$+$10 &  2.6200$+$10 &  4.8370$-$02  \\
    8 &   15 &  3.9426$+$11 &  3.9350$+$11 &  9.6730$-$01  \\
    8 &   21 &  8.6599$+$09 &  8.6610$+$09 &  7.4531$-$03  \\
    8 &   22 &  1.3007$+$11 &  1.3020$+$11 &  1.4923$-$01  \\
    9 &   17 &  1.8435$+$10 &  1.9200$+$10 &  1.3867$-$01  \\
    9 &   18 &  1.8180$+$10 &  1.8940$+$10 &  2.7264$-$01  \\
   10 &   19 &  3.4318$+$10 &  3.5530$+$10 &  5.6204$-$01  \\
   11 &   20 &  4.1273$+$10 &  4.2690$+$10 &  5.0974$-$01  \\
   12 &   18 &  6.2339$+$08 &  6.3650$+$08 &  5.4688$-$03  \\
   12 &   21 &  6.8948$+$10 &  6.8450$+$10 &  8.8628$-$01  \\
   13 &   21 &  4.9245$+$09 &  4.8880$+$09 &  4.2288$-$02  \\
   13 &   22 &  7.3842$+$10 &  7.3320$+$10 &  8.4502$-$01  \\
   14 &   23 &  1.1728$+$11 &  1.1730$+$11 &  1.3448$+$00  \\
\hline   
\end{tabular}

\begin{flushleft}
{\small
GRASP: Present results from the {\sc grasp} code \\
FAC:	Present results from the {\sc fac} code \\
}
\end{flushleft}
\end{table*}


\newpage
\clearpage

\begin{table*}
\begin{flushleft}
{\bf Table D.} Comparison of radiative rates (A- values, s$^{-1}$) for some transitions of S XIV.  ($a{\pm}b \equiv a{\times}$10$^{{\pm}b}$). \\
\end{flushleft}
\begin{tabular}{rrrrrr}

\hline

\hline

$i$ & $j$ & A (GRASP) & A (FAC) & A (BPRM) & f (GRASP)        \\

\hline
    1 &    2 &  1.0163$+$09 &  1.0240$+$09 &  9.6770$+$08 &  2.9937$-$02  \\
    1 &    3 &  1.2392$+$09 &  1.2450$+$09 &  1.2280$+$09 &  6.4183$-$02  \\
    1 &    5 &  8.4139$+$11 &  8.6020$+$11 &  8.6000$+$11 &  1.1717$-$01  \\
    1 &    6 &  8.3110$+$11 &  8.5070$+$11 &  8.4790$+$11 &  2.3084$-$01  \\
    1 &   10 &  3.7633$+$11 &  4.0060$+$11 &  3.5950$+$11 &  2.9899$-$02  \\
    1 &   11 &  3.7301$+$11 &  3.9760$+$11 &  3.5590$+$11 &  5.9222$-$02  \\
    1 &   17 &  1.9527$+$11 &  2.2930$+$11 &  1.8430$+$11 &  1.2527$-$02  \\
    1 &   18 &  1.9376$+$11 &  2.2790$+$11 &  1.8280$+$11 &  2.4852$-$02  \\
    2 &    7 &  2.1329$+$12 &  2.1440$+$12 &  2.1320$+$12 &  6.7297$-$01  \\
    2 &   12 &  6.9776$+$11 &  7.1460$+$11 &  6.9110$+$11 &  1.2268$-$01  \\
    2 &   19 &  3.2220$+$11 &  3.5390$+$11 &  3.1880$+$11 &  4.5375$-$02  \\
    3 &    8 &  2.5470$+$12 &  2.5620$+$12 &  2.5440$+$12 &  6.0807$-$01  \\
    3 &   13 &  8.3118$+$11 &  8.5150$+$11 &  8.2240$+$11 &  1.1037$-$01  \\
    3 &   20 &  3.8323$+$11 &  4.2020$+$11 &  3.7890$+$11 &  4.0736$-$02  \\
    4 &    5 &  1.2993$+$08 &  1.3000$+$08 &  1.1710$+$08 &  5.0332$-$02  \\
    4 &    6 &  1.6019$+$08 &  1.5990$+$08 &  1.5120$+$08 &  1.0816$-$01  \\
    4 &   10 &  1.0777$+$11 &  1.0980$+$11 &  1.0530$+$11 &  1.2767$-$01  \\
    4 &   11 &  1.0621$+$11 &  1.0820$+$11 &  1.0350$+$11 &  2.5082$-$01  \\
    4 &   17 &  5.9998$+$10 &  6.5670$+$10 &  5.8230$+$10 &  3.4481$-$02  \\
    4 &   18 &  5.9322$+$10 &  6.4960$+$10 &  5.7480$+$10 &  6.8104$-$02  \\
    5 &    7 &  7.0108$+$06 &              &  7.3810$+$06 &  2.8520$-$02  \\
    5 &   12 &  2.2339$+$11 &  2.2580$+$11 &  2.2280$+$11 &  5.8063$-$01  \\
    5 &   19 &  1.1025$+$11 &  1.1560$+$11 &  1.1000$+$11 &  1.3607$-$01  \\
    6 &    8 &  5.8918$+$06 &              &  5.7810$+$06 &  2.2803$-$02  \\
    6 &   13 &  2.6807$+$11 &  2.7090$+$11 &  2.6750$+$11 &  5.2634$-$01  \\
    6 &   20 &  1.3193$+$11 &  1.3810$+$11 &  1.3170$+$11 &  1.2277$-$01  \\
    7 &   11 &  1.5633$+$09 &  1.5870$+$09 &  4.2390$+$11 &  2.1774$-$03  \\
    7 &   14 &  4.9561$+$11 &  4.9460$+$11 &  4.9650$+$11 &  1.0148$+$00  \\
    7 &   18 &  6.7010$+$08 &  7.4580$+$08 &  7.8160$+$05 &  4.3100$-$04  \\
    7 &   21 &  1.6361$+$11 &  1.6380$+$11 &  1.6380$+$11 &  1.5674$-$01  \\
    8 &   14 &  3.5336$+$10 &  3.5250$+$10 &  3.5390$+$10 &  4.8366$-$02  \\
    8 &   15 &  5.3027$+$11 &  5.2940$+$11 &  5.3110$+$11 &  9.6720$-$01  \\
    8 &   21 &  1.1645$+$10 &  1.1650$+$10 &  1.1650$+$10 &  7.4508$-$03  \\
    8 &   22 &  1.7493$+$11 &  1.7520$+$11 &  1.7510$+$11 &  1.4921$-$01  \\
    9 &   10 &  3.0289$+$07 &              &  6.3150$+$07 &  6.9479$-$02  \\
    9 &   11 &  3.7493$+$07 &              &  7.7270$+$07 &  1.4942$-$01  \\
    9 &   17 &  2.5083$+$10 &  2.6050$+$10 &  2.3940$+$10 &  1.4138$-$01  \\
    9 &   18 &  2.4689$+$10 &  2.5650$+$10 &  2.3470$+$10 &  2.7729$-$01  \\
   10 &   12 &  2.1583$+$06 &              &  3.6340$+$06 &  5.0597$-$02  \\
   10 &   19 &  4.6223$+$10 &  4.7690$+$10 &  4.5860$+$10 &  5.6353$-$01  \\
   11 &   13 &  1.8066$+$06 &              &  3.0020$+$06 &  4.0397$-$02  \\
   11 &   20 &  5.5610$+$10 &  5.7330$+$10 &  5.5230$+$10 &  5.1171$-$01  \\
   12 &   14 &  1.1951$+$03 &              &  1.5170$+$03 &  2.4526$-$03  \\
   12 &   18 &  8.3155$+$08 &  8.4930$+$08 &  4.4680$+$10 &  5.4151$-$03  \\
   12 &   21 &  9.2746$+$10 &  9.2170$+$10 &  9.2680$+$10 &  8.8604$-$01  \\
   13 &   14 &  7.3981$+$00 &              &  7.0890$+$00 &  5.1660$-$05  \\
   13 &   15 &  4.8212$+$02 &              &  5.5380$+$02 &  1.6869$-$03  \\
   13 &   21 &  6.6242$+$09 &  6.5820$+$09 &  6.6190$+$09 &  4.2291$-$02  \\
   13 &   22 &  9.9322$+$10 &  9.8720$+$10 &  9.9240$+$10 &  8.4495$-$01  \\
   14 &   23 &  1.5776$+$11 &  1.5770$+$11 &  1.5770$+$11 &  1.3447$+$00  \\
   16 &   17 &  9.8151$+$06 &              &  2.4260$+$07 &  8.8162$-$02  \\
   16 &   18 &  1.2172$+$07 &              &  2.9400$+$07 &  1.8966$-$01  \\
   17 &   19 &  7.7673$+$05 &              &  1.4200$+$06 &  7.0329$-$02  \\
   18 &   20 &  6.4914$+$05 &              &  1.1860$+$06 &  5.6122$-$02  \\
   19 &   21 &  6.5279$+$02 &              &  8.2040$+$02 &  4.6843$-$03  \\
   20 &   21 &  4.7351$+$00 &              &  4.5260$+$00 &  1.0403$-$04  \\
   20 &   22 &  2.7679$+$02 &              &  3.1440$+$02 &  3.2756$-$03  \\
   21 &   23 &  8.2320$+$00 &              &  1.0130$+$01 &  6.9141$-$04  \\
\hline   
\end{tabular}

\begin{flushleft}
{\small
GRASP: Present results from the {\sc grasp} code \\
FAC:	Present results from the {\sc fac} code \\
BPRM:  Nahar \cite{sn} \\
}
\end{flushleft}
\end{table*}


\newpage
\clearpage

\begin{table*}
\begin{flushleft}
{\bf Table E.} Comparison of radiative rates (A- values, s$^{-1}$) for some transitions of Cl XV.  ($a{\pm}b \equiv a{\times}$10$^{{\pm}b}$). \\
\end{flushleft}
\begin{tabular}{rrrrr}

\hline

\hline

$i$ & $j$ & A (GRASP) & A (FAC) &  f (GRASP)        \\

\hline
    1 &    2 &  1.0928$+$09 &  1.1000$+$09 &  2.7998$-$02  \\
    1 &    3 &  1.3907$+$09 &  1.3960$+$09 &  6.0922$-$02  \\
    1 &    5 &  1.1134$+$12 &  1.1370$+$12 &  1.1883$-$01  \\
    1 &    6 &  1.0979$+$12 &  1.1220$+$12 &  2.3364$-$01  \\
    1 &   10 &  4.9669$+$11 &  5.2650$+$11 &  3.0176$-$02  \\
    1 &   11 &  4.9178$+$11 &  5.2200$+$11 &  5.9695$-$02  \\
    1 &   17 &  2.5752$+$11 &  2.9890$+$11 &  1.2624$-$02  \\
    1 &   18 &  2.5529$+$11 &  2.9690$+$11 &  2.5018$-$02  \\
    2 &    7 &  2.8087$+$12 &  2.8230$+$12 &  6.7342$-$01  \\
    2 &   12 &  9.1809$+$11 &  9.3880$+$11 &  1.2261$-$01  \\
    2 &   19 &  4.2383$+$11 &  4.6260$+$11 &  4.5331$-$02  \\
    3 &    8 &  3.3516$+$12 &  3.3710$+$12 &  6.0883$-$01  \\
    3 &   13 &  1.0925$+$12 &  1.1180$+$12 &  1.1030$-$01  \\
    3 &   20 &  5.0348$+$11 &  5.4860$+$11 &  4.0684$-$02  \\
    4 &    5 &  1.4013$+$08 &  1.4020$+$08 &  4.7108$-$02  \\
    4 &    6 &  1.8074$+$08 &  1.8040$+$08 &  1.0280$-$01  \\
    4 &   10 &  1.4317$+$11 &  1.4570$+$11 &  1.2969$-$01  \\
    4 &   11 &  1.4082$+$11 &  1.4340$+$11 &  2.5416$-$01  \\
    4 &   17 &  7.9494$+$10 &  8.6470$+$10 &  3.4857$-$02  \\
    4 &   18 &  7.8484$+$10 &  8.5410$+$10 &  6.8735$-$02  \\
    5 &   12 &  2.9456$+$11 &  2.9750$+$11 &  5.8155$-$01  \\
    5 &   19 &  1.4528$+$11 &  1.5180$+$11 &  1.3614$-$01  \\
    6 &   13 &  3.5347$+$11 &  3.5690$+$11 &  5.2772$-$01  \\
    6 &   20 &  1.7378$+$11 &  1.8130$+$11 &  1.2289$-$01  \\
    7 &   11 &  2.0430$+$09 &  2.0740$+$09 &  2.1565$-$03  \\
    7 &   14 &  6.5322$+$11 &  6.5210$+$11 &  1.0146$+$00  \\
    7 &   18 &  8.7602$+$08 &  9.7020$+$08 &  4.2730$-$04  \\
    7 &   21 &  2.1565$+$11 &  2.1590$+$11 &  1.5673$-$01  \\
    8 &   14 &  4.6561$+$10 &  4.6450$+$10 &  4.8361$-$02  \\
    8 &   15 &  6.9876$+$11 &  6.9780$+$11 &  9.6708$-$01  \\
    8 &   21 &  1.5341$+$10 &  1.5340$+$10 &  7.4484$-$03  \\
    8 &   22 &  2.3051$+$11 &  2.3090$+$11 &  1.4919$-$01  \\
    9 &   17 &  3.3387$+$10 &  3.4590$+$10 &  1.4376$-$01  \\
    9 &   18 &  3.2793$+$10 &  3.3990$+$10 &  2.8121$-$01  \\
   10 &   19 &  6.0982$+$10 &  6.2740$+$10 &  5.6467$-$01  \\
   11 &   20 &  7.3392$+$10 &  7.5450$+$10 &  5.1340$-$01  \\
   12 &   18 &  1.0878$+$09 &  1.1110$+$09 &  5.3678$-$03  \\
   12 &   21 &  1.2223$+$11 &  1.2160$+$11 &  8.8577$-$01  \\
   13 &   21 &  8.7300$+$09 &  8.6810$+$09 &  4.2292$-$02  \\
   13 &   22 &  1.3089$+$11 &  1.3020$+$11 &  8.4485$-$01  \\
   14 &   23 &  2.0792$+$11 &  2.0790$+$11 &  1.3445$+$00  \\
\hline   
\end{tabular}

\begin{flushleft}
{\small
GRASP: Present results from the {\sc grasp} code \\
FAC:	Present results from the {\sc fac} code \\
}
\end{flushleft}
\end{table*}


\newpage
\clearpage

\begin{table*}
\begin{flushleft}
{\bf Table F.} Comparison of radiative rates (A- values, s$^{-1}$) for some transitions of Ar XVI.  ($a{\pm}b \equiv a{\times}$10$^{{\pm}b}$). \\
\end{flushleft}
\begin{tabular}{rrrrrr}

\hline

\hline

$i$ & $j$ & A (GRASP) & A (FAC) & A (BPRM) & f (GRASP)        \\

\hline
    1 &    2 &  1.1702$+$09 &  1.1780$+$09 &  1.1130$+$09 &  2.6298$-$02  \\
    1 &    3 &  1.5626$+$09 &  1.5680$+$09 &  1.5650$+$09 &  5.8179$-$02  \\
    1 &    5 &  1.4466$+$12 &  1.4750$+$12 &  1.4820$+$12 &  1.2029$-$01  \\
    1 &    6 &  1.4241$+$12 &  1.4540$+$12 &  1.4560$+$12 &  2.3601$-$01  \\
    1 &   10 &  6.4389$+$11 &  6.7990$+$11 &  5.9760$+$11 &  3.0416$-$02  \\
    1 &   11 &  6.3677$+$11 &  6.7350$+$11 &  5.9010$+$11 &  6.0093$-$02  \\
    1 &   17 &  3.3359$+$11 &  3.8350$+$11 &  3.1310$+$11 &  1.2707$-$02  \\
    1 &   18 &  3.3040$+$11 &  3.8050$+$11 &  3.0980$+$11 &  2.5158$-$02  \\
    2 &    7 &  3.6340$+$12 &  3.6520$+$12 &  3.6350$+$12 &  6.7371$-$01  \\
    2 &   12 &  1.1871$+$12 &  1.2120$+$12 &  1.1770$+$12 &  1.2254$-$01  \\
    2 &   19 &  5.4788$+$11 &  5.9480$+$11 &  5.4080$+$11 &  4.5292$-$02  \\
    3 &    8 &  4.3331$+$12 &  4.3580$+$12 &  4.3300$+$12 &  6.0947$-$01  \\
    3 &   13 &  1.4109$+$12 &  1.4420$+$12 &  1.3970$+$12 &  1.1024$-$01  \\
    3 &   20 &  6.4999$+$11 &  7.0440$+$11 &  6.4120$+$11 &  4.0636$-$02  \\
    4 &    5 &  1.5046$+$08 &  1.5050$+$08 &  1.3420$+$08 &  4.4278$-$02  \\
    4 &    6 &  2.0413$+$08 &  2.0370$+$08 &  1.9350$+$08 &  9.8303$-$02  \\
    4 &   10 &  1.8665$+$11 &  1.8980$+$11 &  1.7960$+$11 &  1.3147$-$01  \\
    4 &   11 &  1.8322$+$11 &  1.8640$+$11 &  1.7570$+$11 &  2.5699$-$01  \\
    4 &   17 &  1.0340$+$11 &  1.1190$+$11 &  1.0030$+$11 &  3.5185$-$02  \\
    4 &   18 &  1.0193$+$11 &  1.1030$+$11 &  9.8680$+$10 &  6.9264$-$02  \\
    5 &    7 &  1.0465$+$07 &		   &  1.1110$+$07 &  2.7258$-$02  \\
    5 &   12 &  3.8150$+$11 &  3.8500$+$11 &  3.8070$+$11 &  5.8222$-$01  \\
    5 &   19 &  1.8807$+$11 &  1.9590$+$11 &  1.8740$+$11 &  1.3619$-$01  \\
    6 &    8 &  7.6548$+$06 &		   &  7.4420$+$06 &  2.0810$-$02  \\
    6 &   13 &  4.5780$+$11 &  4.6200$+$11 &  4.5700$+$11 &  5.2892$-$01  \\
    6 &   20 &  2.2487$+$11 &  2.3400$+$11 &  2.2410$+$11 &  1.2298$-$01  \\
    7 &   11 &  2.6251$+$09 &  2.6650$+$09 &  7.2150$+$11 &  2.1379$-$03  \\
    7 &   14 &  8.4576$+$11 &  8.4460$+$11 &  8.4750$+$11 &  1.0144$+$00  \\
    7 &   18 &  1.1260$+$09 &  1.2410$+$09 &  9.1910$+$05 &  4.2401$-$04  \\
    7 &   21 &  2.7924$+$11 &  2.7960$+$11 &  2.7960$+$11 &  1.5672$-$01  \\
    8 &   14 &  6.0268$+$10 &  6.0140$+$10 &  6.0380$+$10 &  4.8356$-$02  \\
    8 &   15 &  9.0453$+$11 &  9.0360$+$11 &  9.0620$+$11 &  9.6696$-$01  \\
    8 &   21 &  1.9853$+$10 &  1.9850$+$10 &  1.9870$+$10 &  7.4459$-$03  \\
    8 &   22 &  2.9838$+$11 &  2.9890$+$11 &  2.9860$+$11 &  1.4917$-$01  \\
    9 &   10 &  3.5147$+$07 &		   &  1.1740$+$08 &  6.1153$-$02  \\
    9 &   11 &  4.7952$+$07 &		   &  1.5080$+$08 &  1.3591$-$01  \\
    9 &   17 &  4.3601$+$10 &  4.5080$+$10 &  4.1140$+$10 &  1.4587$-$01  \\
    9 &   18 &  4.2729$+$10 &  4.4190$+$10 &  4.0100$+$10 &  2.8452$-$01  \\
   10 &   12 &  3.2264$+$06 &		   &  5.5310$+$06 &  4.8391$-$02  \\
   10 &   19 &  7.9015$+$10 &  8.1100$+$10 &  7.8200$+$10 &  5.6550$-$01  \\
   11 &   13 &  2.3471$+$06 &		   &  3.9930$+$06 &  3.6873$-$02  \\
   11 &   20 &  9.5133$+$10 &  9.7570$+$10 &  9.4270$+$10 &  5.1486$-$01  \\
   12 &   14 &  2.9725$+$03 &		   &  3.6010$+$03 &  2.7806$-$03  \\
   12 &   18 &  1.3990$+$09 &  1.4290$+$09 &  7.6370$+$10 &  5.3256$-$03  \\
   12 &   21 &  1.5825$+$11 &  1.5750$+$11 &  1.5810$+$11 &  8.8548$-$01  \\
   13 &   14 &  9.6760$+$00 &		   &  7.7610$+$00 &  4.7266$-$05  \\
   13 &   15 &  9.9815$+$02 &		   &  1.0680$+$03 &  1.7991$-$03  \\
   13 &   21 &  1.1302$+$10 &  1.1250$+$10 &  1.1290$+$10 &  4.2294$-$02  \\
   13 &   22 &  1.6944$+$11 &  1.6870$+$11 &  1.6930$+$11 &  8.4474$-$01  \\
   14 &   23 &  2.6918$+$11 &  2.6920$+$11 &  2.6890$+$11 &  1.3443$+$00  \\
   16 &   17 &  1.1402$+$07 &		   &  3.5650$+$07 &  7.7615$-$02  \\
   16 &   18 &  1.5597$+$07 &		   &  4.6170$+$07 &  1.7258$-$01  \\
   17 &   19 &  1.1619$+$06 &		   &  2.5510$+$06 &  6.7286$-$02  \\
   18 &   20 &  8.4341$+$05 &		   &  1.9130$+$06 &  5.1233$-$02  \\
   19 &   21 &  1.5827$+$03 &		   &  1.9030$+$03 &  5.2658$-$03  \\
   20 &   21 &  6.1818$+$00 &		   &  5.0000$+$00 &  9.5131$-$05  \\
   20 &   22 &  5.5839$+$02 &		   &  5.9250$+$02 &  3.4635$-$03  \\
   21 &   23 &  3.0548$+$01 &		   &  3.5240$+$01 &  8.9579$-$04  \\
\hline   
\end{tabular}

\begin{flushleft}
{\small
GRASP: Present results from the {\sc grasp} code \\
FAC:	Present results from the {\sc fac} code \\
BPRM:  Nahar \cite{sn} \\
}
\end{flushleft}
\end{table*}


\newpage
\clearpage

\begin{table*}
\begin{flushleft}
{\bf Table G.} Comparison of radiative rates (A- values, s$^{-1}$) for some transitions of K XVII.  ($a{\pm}b \equiv a{\times}$10$^{{\pm}b}$). \\
\end{flushleft}
\begin{tabular}{rrrrr}

\hline

\hline

$i$ & $j$ & A (GRASP) & A (FAC) &  f (GRASP)        \\

\hline
    1 &    2 &  1.2486$+$09 &  1.2560$+$09 &  2.4795$-$02  \\
    1 &    3 &  1.7587$+$09 &  1.7640$+$09 &  5.5870$-$02  \\
    1 &    5 &  1.8498$+$12 &  1.8840$+$12 &  1.2160$-$01  \\
    1 &    6 &  1.8176$+$12 &  1.8540$+$12 &  2.3801$-$01  \\
    1 &   10 &  8.2167$+$11 &  8.6470$+$11 &  3.0628$-$02  \\
    1 &   11 &  8.1159$+$11 &  8.5570$+$11 &  6.0426$-$02  \\
    1 &   17 &  4.2543$+$11 &  4.8480$+$11 &  1.2780$-$02  \\
    1 &   18 &  4.2093$+$11 &  4.8080$+$11 &  2.5274$-$02  \\
    2 &    7 &  4.6295$+$12 &  4.6520$+$12 &  6.7387$-$01  \\
    2 &   12 &  1.5114$+$12 &  1.5420$+$12 &  1.2248$-$01  \\
    2 &   19 &  6.9747$+$11 &  7.5350$+$11 &  4.5256$-$02  \\
    3 &    8 &  5.5156$+$12 &  5.5470$+$12 &  6.1001$-$01  \\
    3 &   13 &  1.7943$+$12 &  1.8310$+$12 &  1.1017$-$01  \\
    3 &   20 &  8.2631$+$11 &  8.9110$+$11 &  4.0591$-$02  \\
    4 &    6 &  2.3092$+$08 &  2.3030$+$08 &  9.4531$-$02  \\
    4 &   10 &  2.3939$+$11 &  2.4320$+$11 &  1.3306$-$01  \\
    4 &   11 &  2.3447$+$11 &  2.3830$+$11 &  2.5937$-$01  \\
    4 &   17 &  1.3234$+$11 &  1.4250$+$11 &  3.5476$-$02  \\
    4 &   18 &  1.3025$+$11 &  1.4030$+$11 &  6.9707$-$02  \\
    5 &   12 &  4.8641$+$11 &  4.9060$+$11 &  5.8267$-$01  \\
    5 &   19 &  2.3969$+$11 &  2.4910$+$11 &  1.3622$-$01  \\
    6 &   13 &  5.8367$+$11 &  5.8870$+$11 &  5.2994$-$01  \\
    6 &   20 &  2.8648$+$11 &  2.9740$+$11 &  1.2306$-$01  \\
    7 &   11 &  3.3232$+$09 &  3.3760$+$09 &  2.1212$-$03  \\
    7 &   14 &  1.0781$+$12 &  1.0770$+$12 &  1.0141$+$00  \\
    7 &   18 &  1.4259$+$09 &  1.5670$+$09 &  4.2107$-$04  \\
    7 &   21 &  3.5598$+$11 &  3.5640$+$11 &  1.5671$-$01  \\
    8 &   14 &  7.6797$+$10 &  7.6640$+$10 &  4.8350$-$02  \\
    8 &   15 &  1.1527$+$12 &  1.1520$+$12 &  9.6682$-$01  \\
    8 &   21 &  2.5293$+$10 &  2.5290$+$10 &  7.4432$-$03  \\
    8 &   22 &  3.8024$+$11 &  3.8090$+$11 &  1.4915$-$01  \\
    9 &   17 &  5.6001$+$10 &  5.7790$+$10 &  1.4774$-$01  \\
    9 &   18 &  5.4751$+$10 &  5.6520$+$10 &  2.8730$-$01  \\
   10 &   19 &  1.0077$+$11 &  1.0320$+$11 &  5.6605$-$01  \\
   11 &   20 &  1.2138$+$11 &  1.2420$+$11 &  5.1612$-$01  \\
   12 &   18 &  1.7725$+$09 &  1.8100$+$09 &  5.2879$-$03  \\
   12 &   21 &  2.0169$+$11 &  2.0090$+$11 &  8.8515$-$01  \\
   13 &   21 &  1.4405$+$10 &  1.4340$+$10 &  4.2295$-$02  \\
   13 &   22 &  2.1594$+$11 &  2.1510$+$11 &  8.4462$-$01  \\
   14 &   23 &  3.4309$+$11 &  3.4310$+$11 &  1.3441$+$00  \\
\hline   
\end{tabular}

\begin{flushleft}
{\small
GRASP: Present results from the {\sc grasp} code \\
FAC:	Present results from the {\sc fac} code \\
}
\end{flushleft}
\end{table*}


\newpage
\clearpage

\begin{table*}
\begin{flushleft}
{\bf Table H.} Comparison of radiative rates (A- values, s$^{-1}$) for some transitions of Ca XVIII.  ($a{\pm}b \equiv a{\times}$10$^{{\pm}b}$). \\
\end{flushleft}
\begin{tabular}{rrrrrr}

\hline

\hline

$i$ & $j$ & A (GRASP) & A (FAC) & A (BPRM) & f (GRASP)        \\

\hline
    1 &    2 &  1.3281$+$09 &  1.3360$+$09 &  1.2550$+$09 &  2.3456$-$02  \\
    1 &    3 &  1.9838$+$09 &  1.9890$+$09 &  2.0010$+$09 &  5.3933$-$02  \\
    1 &    5 &  2.3322$+$12 &  2.3720$+$12 &  2.3970$+$12 &  1.2276$-$01  \\
    1 &    6 &  2.2872$+$12 &  2.3310$+$12 &  2.3460$+$12 &  2.3969$-$01  \\
    1 &   10 &  1.0341$+$12 &  1.0850$+$12 &  9.3850$+$11 &  3.0814$-$02  \\
    1 &   11 &  1.0200$+$12 &  1.0730$+$12 &  9.2410$+$11 &  6.0705$-$02  \\
    1 &   17 &  5.3508$+$11 &  6.0520$+$11 &  4.9450$+$11 &  1.2843$-$02  \\
    1 &   18 &  5.2886$+$11 &  5.9960$+$11 &  4.8820$+$11 &  2.5371$-$02  \\
    2 &    7 &  5.8174$+$12 &  5.8440$+$12 &  5.8240$+$12 &  6.7389$-$01  \\
    2 &   12 &  1.8984$+$12 &  1.9340$+$12 &  1.8790$+$12 &  1.2241$-$01  \\
    2 &   19 &  8.7595$+$11 &  9.4250$+$11 &  8.6310$+$11 &  4.5224$-$02  \\
    3 &    8 &  6.9247$+$12 &  6.9640$+$12 &  6.9240$+$12 &  6.1046$-$01  \\
    3 &   13 &  2.2508$+$12 &  2.2950$+$12 &  2.2250$+$12 &  1.1011$-$01  \\
    3 &   20 &  1.0362$+$12 &  1.1130$+$12 &  1.0200$+$12 &  4.0548$-$02  \\
    4 &    5 &  1.7155$+$08 &		   &  1.5030$+$08 &  3.9547$-$02  \\
    4 &    6 &  2.6181$+$08 &  2.6110$+$08 &  2.4820$+$08 &  9.1380$-$02  \\
    4 &   10 &  3.0261$+$11 &  3.0720$+$11 &  2.8760$+$11 &  1.3447$-$01  \\
    4 &   11 &  2.9571$+$11 &  3.0040$+$11 &  2.7990$+$11 &  2.6137$-$01  \\
    4 &   17 &  1.6699$+$11 &  1.7900$+$11 &  1.6070$+$11 &  3.5733$-$02  \\
    4 &   18 &  1.6407$+$11 &  1.7600$+$11 &  1.5750$+$11 &  7.0078$-$02  \\
    5 &    7 &  1.5776$+$07 &		   &  1.6980$+$07 &  2.6693$-$02  \\
    5 &   12 &  6.1159$+$11 &  6.1650$+$11 &  6.0960$+$11 &  5.8291$-$01  \\
    5 &   19 &  3.0131$+$11 &  3.1250$+$11 &  3.0000$+$11 &  1.3623$-$01  \\
    6 &    8 &  9.8469$+$06 &		   &  9.5370$+$06 &  1.9332$-$02  \\
    6 &   13 &  7.3388$+$11 &  7.3990$+$11 &  7.3190$+$11 &  5.3083$-$01  \\
    6 &   20 &  3.5996$+$11 &  3.7290$+$11 &  3.5850$+$11 &  1.2313$-$01  \\
    7 &   11 &  4.1519$+$09 &  4.2170$+$09 &  1.1540$+$12 &  2.1062$-$03  \\
    7 &   14 &  1.3553$+$12 &  1.3540$+$12 &  1.3590$+$12 &  1.0138$+$00  \\
    7 &   18 &  1.7820$+$09 &  1.9510$+$09 &  1.0540$+$06 &  4.1843$-$04  \\
    7 &   21 &  4.4756$+$11 &  4.4810$+$11 &  4.4830$+$11 &  1.5670$-$01  \\
    8 &   14 &  9.6512$+$10 &  9.6320$+$10 &  9.6720$+$10 &  4.8343$-$02  \\
    8 &   15 &  1.4487$+$12 &  1.4480$+$12 &  1.4520$+$12 &  9.6667$-$01  \\
    8 &   21 &  3.1780$+$10 &  3.1770$+$10 &  3.1810$+$10 &  7.4404$-$03  \\
    8 &   22 &  4.7789$+$11 &  4.7880$+$11 &  4.7840$+$11 &  1.4913$-$01  \\
    9 &   10 &  4.0133$+$07 &		   &  1.9780$+$08 &  5.4643$-$02  \\
    9 &   11 &  6.1715$+$07 &		   &  2.6850$+$08 &  1.2645$-$01  \\
    9 &   17 &  7.0881$+$10 &  7.3020$+$10 &  6.6100$+$10 &  1.4941$-$01  \\
    9 &   18 &  6.9125$+$10 &  7.1240$+$10 &  6.4000$+$10 &  2.8963$-$01  \\
   10 &   12 &  4.8748$+$06 &		   &  1.0070$+$07 &  4.7432$-$02  \\
   10 &   19 &  1.2673$+$11 &  1.2960$+$11 &  1.2500$+$11 &  5.6635$-$01  \\
   11 &   13 &  3.0216$+$06 &		   &  6.4710$+$06 &  3.4268$-$02  \\
   11 &   20 &  1.5271$+$11 &  1.5600$+$11 &  1.5090$+$11 &  5.1720$-$01  \\
   12 &   14 &  7.1069$+$03 &		   &  8.3380$+$03 &  3.1772$-$03  \\
   12 &   18 &  2.2161$+$09 &  2.2630$+$09 &  1.2240$+$11 &  5.2539$-$03  \\
   12 &   21 &  2.5353$+$11 &  2.5260$+$11 &  2.5320$+$11 &  8.8479$-$01  \\
   13 &   14 &  1.2050$+$01 &		   &  7.9590$+$00 &  4.3448$-$05  \\
   13 &   15 &  2.0351$+$03 &		   &  2.0670$+$03 &  1.9495$-$03  \\
   13 &   21 &  1.8107$+$10 &  1.8040$+$10 &  1.8090$+$10 &  4.2295$-$02  \\
   13 &   22 &  2.7142$+$11 &  2.7050$+$11 &  2.7110$+$11 &  8.4448$-$01  \\
   14 &   23 &  4.3126$+$11 &  4.3130$+$11 &  4.3080$+$11 &  1.3439$+$00  \\
   16 &   17 &  1.3029$+$07 &		   &  5.6360$+$07 &  6.9369$-$02  \\
   16 &   18 &  2.0108$+$07 &		   &  7.7830$+$07 &  1.6064$-$01  \\
   17 &   19 &  1.7574$+$06 &		   &  4.8060$+$06 &  6.5982$-$02  \\
   18 &   20 &  1.0863$+$06 &		   &  3.2630$+$06 &  4.7624$-$02  \\
   19 &   21 &  3.7064$+$03 &		   &  4.3250$+$03 &  5.9757$-$03  \\
   20 &   21 &  7.6892$+$00 &		   &  5.1810$+$00 &  8.7415$-$05  \\
   20 &   22 &  1.1122$+$03 &		   &  1.1210$+$03 &  3.7239$-$03  \\
   21 &   23 &  9.7663$+$01 &		   &  1.0860$+$02 &  1.1278$-$03  \\
\hline   
\end{tabular}

\begin{flushleft}
{\small
GRASP: Present results from the {\sc grasp} code \\
FAC:	Present results from the {\sc fac} code \\
BPRM:  Nahar \cite{sn} \\
}
\end{flushleft}
\end{table*}

\end{document}